\documentclass[%
 reprint,
 amsmath,amssymb,
 aps,
floatfix,
]{revtex4-1}

\usepackage{graphicx}% Include figure files
\usepackage{dcolumn}% Align table columns on decimal point
\usepackage{bm}% bold math
\usepackage{units}
\usepackage{float}
\usepackage{microtype}
\usepackage{xcolor}
\begin{document}

%\preprint{APS/123-QED}

\title{Influence of hydrogenation on the vibrational density of states of magnetocaloric $\mathrm{LaFe}_\mathrm{11.4}\mathrm{Si}_\mathrm{1.6}\mathrm{H}_{1.6}$}% Force line breaks with \\

\author{A. Terwey$^{1}$}
\author{M. E. Gruner$^{1}$}
\author{W. Keune$^{1}$}
\author{J. Landers$^{1}$}
\author{S. Salamon$^{1}$}
\author{B. Eggert$^{1}$}
\author{K. Ollefs$^{1}$}
\author{V. Brab\"{a}nder$^{2}$}
\author{I. Radulov$^{2}$}
\author{K. Skokov$^{2}$}
\author{T. Faske$^{2}$}
\author{M. Y. Hu$^{3}$}
\author{J. Zhao$^{3}$}
\author{E. E. Alp$^{3}$}
\author{C. Giacobbe$^{4}$}
\author{O. Gutfleisch$^{2}$}
\author{H. Wende$^{1}$}
\email{heiko.wende@uni-due.de}
\affiliation{%
 $^{1}$Faculty of Physics and Center for Nanointegration Duisburg-Essen (CENIDE), University of Duisburg-Essen, 47057 Duisburg, Germany\\
 $^{2}$Materials Science, TU Darmstadt, 64287 Darmstadt, Germany\\
 $^{3}$Advanced Photon Source (APS), Argonne National Laboratory, Lemont, IL 60439, USA\\
 $^{4}$European Synchrotron Radiation Facility (ESRF), 38000 Grenoble, France\\
}%

\date{\today}
\begin{abstract}

  We report on the impact of magnetoelastic coupling on the magnetocaloric properties of
  $\mathrm{LaFe}_\mathrm{11.4}\mathrm{Si}_\mathrm{1.6}\mathrm{H}_{1.6}$
  in terms of the vibrational (phonon) density of states (VDOS), which we determined with  $^{57}$Fe nuclear resonant
  inelastic X-ray scattering (NRIXS) measurements and with density-functional-theory (DFT)-based
  first-principles calculations in the ferromagnetic (FM) low-temperature and paramagnetic (PM) high-temperature phase.
  In experiments and calculations, we observe pronounced differences in the shape of the Fe-partial VDOS between
  non-hydrogenated and hydrogenated samples. This shows that hydrogen does not only shift the temperature of the
  first-order phase transition, but also affects the elastic response of the Fe-subsystem significantly.
  In turn, the anomalous redshift of the Fe VDOS, observed by going to the low-volume
  PM phase, survives hydrogenation. 
  As a consequence, the change in the Fe specific vibrational entropy $\Delta S_\mathrm{lat}$ across the phase transition has 
  the same sign as the magnetic and electronic
  contribution. DFT calculations show that the same mechanism,
  which is a consequence of the itinerant electron metamagnetism associated with the Fe subsystem,
  is effective in both the hydrogenated and the hydrogen-free compounds.
  Although reduced by \unit[50]{\%} as compared to the hydrogen-free system, the measured change $\Delta S_\mathrm{lat}$ of $\unit[(3.2\pm1.9)]{\frac{J}{kg K}}$ across the FM to PM transition contributes with $\sim\unit[35]{\%}$ significantly and cooperatively to the total isothermal entropy change $\Delta S_\mathrm{iso}$. Hydrogenation is observed to induce an overall blueshift of the Fe-VDOS with respect to the H-free compound; this effect, together with the enhanced Debye temperature observed, is a fingerprint of the hardening of the Fe sublattice by hydrogen incorporation. In addition, the mean Debye velocity of sound of $\mathrm{LaFe}_\mathrm{11.4}\mathrm{Si}_\mathrm{1.6}\mathrm{H}_{1.6}$ was determined from the NRIXS and the DFT data.
%\begin{description}
%\end{description}
\end{abstract}
\maketitle

%\tableofcontents
\section{\label{sec:intro}Introduction}

In the search for an environmentally-friendly
alternative to the conventional gas-compressor refrigeration,
solid state cooling concepts offer an energy efficient perspective
\cite{cn:Faehler12FerroicCooling,cn:Moya14NMAT}.
First-order magnetocaloric materials are considered as one
important class of systems for this
purpose \cite{cn:Pecharsky97PRL,cn:Gschneidner05,cn:Dung11AEM,cn:Boeije16CM,cn:Waske18}.
%attachment A 
Moreover, magnetocaloric materials are considered for local heating and cooling inside the human body \cite{Tishin2016}.
%end attachment A
These materials are characterized by a significant adiabatic temperature change
$|\Delta T_\mathrm{ad}|$ induced by magnetic fields and a large isothermal entropy change
$|\Delta S_\mathrm{iso}|$ at a magnetostructural phase transition at the phase
transition temperature, $T_\mathrm{tr}$.

Among the materials of current interest with a large magnetocaloric effect
and ability to be tailored to possible user applications are
$\mathrm{LaFe}_\mathrm{13-x}\mathrm{Si}_\mathrm{x}$ based compounds
\cite{Gutfleisch2016,Sandeman_2012,Lyubina_2010,Fujieda_2002,fujieda_2001,Liu_2011,Hu2003}.
Their isostructural first-order phase transition at $T_\mathrm{tr}$ is accompanied by a drastic volume decrease with narrow hysteresis \cite{Hu_2001,Kuzmin_2007}, which is associated with an itinerant electron metamagnetic transition (IEM) \cite{spichkin,cn:Gschneidner05,gutfleisch,F_hler_2011,Fujita_1999}. For Si contents of $1.2\leq x \leq 2.5$, $\mathrm{LaFe}_\mathrm{13-x}\mathrm{Si}_\mathrm{x}$ based compounds tend to crystallize in a cubic $\mathrm{NaZn}_{13}$ ($Fm\overline{3}c$) structure (1:13 phase).
The prototype structure is usually represented by a 112 atom unit cell with cubic (cartesian) axes containing two non-equivalent Zn (Fe) sites. Here, we refer with $\mathrm{Fe}_{\mathrm{II}}$ to the 96-fold (96i) sites, which exhibit a lower local symmetry, while $\mathrm{Fe}_{\mathrm{I}}$ corresponds to the highly symmetrical eight-fold (8b) Wyckoff positions. It is widely assumed that Si occupies the (96i) sites randomly, shared with $\mathrm{Fe}_{\mathrm{II}}$ \cite{Hamdeh_2004,Rosca_2010}. The transition temperature in those compounds increases with Si content and turns the phase transition from first to second order, with a reduction in both $|\Delta T_\mathrm{ad}|$ and $|\Delta S_\mathrm{iso}|$ \cite{Jia_2006,Law2018}. To avoid this, the incorporation of hydrogen into these compounds leads to the occupation of the interstitial lattice sites (24d) by hydrogen, whilst retaining a first order phase transition \cite{cn:Fujieda08,Rosca_2010,Krautz_2014}.
Occupation of all (24d) by hydrogen corresponds to $y=3$ H per formula unit. However, in experiment the (24d) sites are never fully occupied and the hydrogen content in $\mathrm{LaFe}_\mathrm{13-x}\mathrm{Si}_\mathrm{x}\mathrm{H}_\mathrm{y}$ does not exceed a value of $y=1.8$ depending on the composition \cite{Wang2009,Phejar2016}.
This is discussed as a result of the non-bonding or even repulsive character of the Si-H interaction \cite{Rosca_2010,Hai2018}. A similar behavior is observed for other hydrogenated compounds than $\mathrm{La(FeSi)}_\mathrm{13}\mathrm{H}_\mathrm{y}$ containing transition-metals and main group elements \cite{Rundqvist1984}.\\

\begin{figure}
    \centering
    \includegraphics[width=0.45\textwidth]{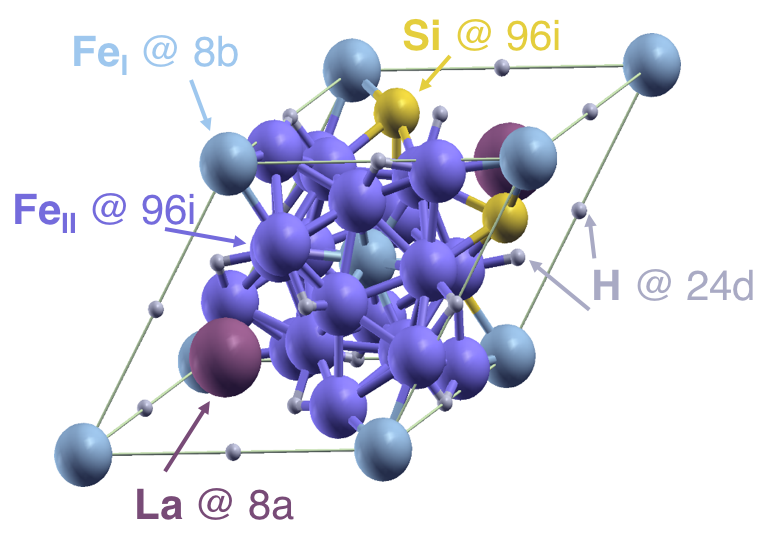}
    \caption{Primitive cell with fcc base for hydrogenated $\mathrm{La}\mathrm{Fe}_{11.5}\mathrm{Si}_{1.5}\mathrm{H}_\mathrm{3}$ consisting of 34 atoms. The colour code for the atoms is as follows: La at (8a) (brown), Fe$_\mathrm{I}$ at (8b) (light blue), Fe$_\mathrm{II}$ at (96i) (dark blue), Si at (96i) (yellow) and H at (24d) (gray)}
    \label{fig:LaFeSiH_cell}
\end{figure}

Fig.\ \ref{fig:LaFeSiH_cell} depicts the primitive cell with a fcc base containing 34 atoms and hydrogen on the interstitial (24d) lattice sites. Interstitial hydrogenation can provide a first order phase transition while retaining good magnetocaloric performance and maintaining a large adiabatic temperature change $|\Delta T_\mathrm{ad}|$ and entropy change $|\Delta S_\mathrm{iso}|$ \cite{Fujita2003,Lyubina2008,Krautz2012,Barcza_2011,Wang_2003}. Hydrogenation and additional incorporation of Mn into these compounds gives the opportunity to specifically tailor the transition to temperatures as needed, covering a broad temperature range (\unit[135-345]{K}) \cite{Krautz_2014} without changing the lattice symmetry \cite{Krautz_2014, Baumfeld_2014} as H increases and Mn decreases the transition temperature drastically. This makes hydrogenated $\mathrm{LaFe}_\mathrm{13-x}\mathrm{Si}_\mathrm{x}$ based compounds promising materials for room temperature cooling applications. 

The total isothermal entropy change $\Delta S_\mathrm{iso}$
is often decomposed into independent contributions from the elemental constituents or the relevant degrees of freedom \cite{spichkin,cn:Fultz10,cn:Scheibel18}:
\begin{equation}
  \Delta S_\mathrm{iso}=\Delta S_\mathrm{mag}+\Delta S_\mathrm{lat}+\Delta S_\mathrm{el}\,.
  \label{eq:Siso}
\end{equation}
The terms on the right hand side refer to the
magnetic, lattice (vibrational) and electronic degrees of freedom, respectively.
One has to bear in mind, that these contributions cannot be separated in a strict sense, since cross-coupling  between the different degrees of freedom -- such as
magnetoelastic or electron-phonon coupling -- must be expected
\cite{spichkin,Gruner_2015,Piazzi2016, basso2017, cn:Landers18, cn:Gercsi18}.
Nevertheless, the simple decomposition, Eq.\ (\ref{eq:Siso}),
still gives a useful indication about the extent
to which a particular set of degrees of freedom contributes to the magnetocaloric
performance of a material.
    
In this work we will illustrate the effect of hydrogenation on the lattice dynamics and thermodynamic properties as well as the structure of the vibrational density of states (VDOS) of $\mathrm{LaFe}_\mathrm{13-x}\mathrm{Si}_\mathrm{x}$ based compounds by means of temperature dependent $^{57}$Fe nuclear resonant inelastic X-ray scattering (NRIXS) and density functional theory (DFT) calculations. NRIXS is directly sensitive to the Fe specific lattice dynamics only \cite{Singwi_1960,Seto_1995,Sturhahn_1995,Chumakov1995}, while DFT provides access to the
contributions from all elements.
The measurement of the phonon excitation probability provides direct access to the $^{57}$Fe-partial vibrational density of states (VDOS) $g(E)$ and the vibrational entropy
$S_\mathrm{lat}$.

%Attachment B
In our present work, we tackle the following open fundamental questions for hydrogenated magnetocaloric La(FeSi)$_{13}$H compounds:
  (i) What is the impact of interstitial hydrogen atoms on the vibrational (phonon) density of states, the latter being a basic property for the understanding of vibrational thermodynamics, such as the vibrational (lattice) entropy $\Delta S_\mathrm{lat}$?
  (ii) How does the hydrogenation affect
the jump of $\Delta S_\mathrm{lat}$ in the temperature dependence of the vibrational entropy observed previously at the FM-to-PM transition in the non-hydrogenated parent material \cite{Gruner_2015,cn:Landers18}?
As suggested earlier \cite{cn:Gercsi18}, corresponding investigations may provide a significant contribution to
quantify the role of electron coupling to the lattice degrees of freedom as a function of hydrogenation.
(iii) Is the magnetostructural first-order FM-to-PM phase transition in the hydrogenated material reflected in the $T$-dependence of the average velocity of sound, $\langle v_D \rangle$, as determined from the NRIXS data? In the literature, this method has been applied to the magnetocaloric (Mn,Fe)$_{1.95}$(P,Si) compound, and $\langle v_D \rangle$ was shown to be larger in the FM state than in the PM state \cite{Bessas2018}.

%end attachment B

\section{\label{sec:exp}Methods}

Experiments have been performed on polycrystalline powder samples (particle size $\leq \unit[100]{\mu m}$) with a nominal composition of $\mathrm{La}\mathrm{Fe}_{11.4}\mathrm{Si}_{1.6}\mathrm{H}_\mathrm{y}$ and $\mathrm{y}\approx1.6$. For better data quality in the NRIXS measurements, the samples are enriched to 30$\%$ in the $^{57}$Fe isotope.
The samples were prepared at the TU Darmstadt by arc melting in Ar atmosphere and subsequent annealing at $\unit[1373]{K}$ for 7 days in an Ar-filled quartz tube followed by quenching in water. The hydrogenation was done by heating the sample in a furnace at $\unit[0.9]{bar}$ $\mathrm{H_2}$ atmosphere at $\unit[723]{K}$ for one hour according to the procedures described in Ref. \cite{Liu_2011,Krautz_2014}.

For the experiments, the ingots were crushed and ground into a powder, as in our former work \cite{Gruner_2015,cn:Landers18}. High resolution powder X-ray diffraction (HR-PXRD) measurements, performed at beamline ID22 of the ESRF, revealed the 1:13 phase with only $\alpha$-Fe as a secondary phase (see supplement in Ref.\ \cite{supp}). The powder prepared for the experiments lead to two samples, A and B, with slightly different residual secondary phase contents. Pre-characterization has been performed via temperature dependent vibrating sample magnetometry (VSM) using a Quantum Design PPMS DynaCool to determine the transition temperature and thermal hysteresis of the compounds. 
%$^{57}$Fe transmission M\"ossbauer spectroscopy and 
Magnetization measurements revealed a first-order IEM transition from a ferro- (FM) to paramagnetic (PM) state at $T_\mathrm{tr}=\unit[329]{K}$ in an applied field of $\mu_0H=\unit[10]{mT}$ for the hydrogenated compounds (see Fig.~S~1 in Ref. \cite{supp}). Their thermal hysteresis obtained by magnetometry is narrow, with a width of only $\unit[3]{K}$. Field dependent magnetometry up to $\unit[9]{T}$ revealed a very good sample quality with a very small $\alpha$-Fe content (secondary phase) of $\unit[1.91]{\%}$ (sample A) and  $\unit[4.47]{\%}$ (sample B). 
%attachement C
Furthermore, we have performed temperature dependent PXRD measurements on an independently prepared hydrogenated $\mathrm{La}\mathrm{Fe}_{11.4}\mathrm{Si}_{1.6}\mathrm{H}_\mathrm{1.6}$  powder sample of natural isotopic composition (sample C) in order to determine the $T$-dependence of the mass density, $\rho$, from the lattice parameter. Details are given in Ref.~\cite{supp}.
%end attachment C

To gather information on the lattice dynamics and to obtain the VDOS, $^{57}$Fe NRIXS \cite{Seto_1995,Sturhahn_1995,Chumakov1995,chumakov1999,Sturhahn_1999} measurements have been performed at Sector 3-ID at the Advanced Photon Source, Argonne National Laboratory. The energy of the incident X-ray beam was tuned around the nuclear resonance energy of $E_0=\unit[14.412]{keV}$ of $^{57}$Fe. The energetic bandwidth of the X-ray beam is reduced with the use of a high-resolution silicon crystal monochromator to a value of \unit[1]{meV} \cite{Toellner_2000}. After passing through a toroidal mirror, the collimated beam was focused onto the sample at grazing incidence relative to the flat sample surface. An avalanche photo diode (APD) detector with timing electronics \cite{Sturhahn2004} was used to detect the $\unit[6.4]{keV}$ fluorescence X-ray signal emitted only with the delayed nuclear resonant scattering events and the $\unit[14.4]{keV}$ fluorescence of the nuclear resonance. For the low temperature data the sample was placed in a closed cycle cryostat under a dome shaped Be window. For the data above room temperature the samples were mounted on a custom built heater at atmospheric pressure. For the NRIXS measurements the powder samples were embedded  in epoxy resin on a Cu plate providing a macroscopically flat sample surface. The experiments were performed in zero external magnetic field as well as with an applied field of $\unit[1.1]{T}$ by permanent magnets. 
Multiple temperature points were taken across $T_\mathrm{tr}$ and in the sample's well defined FM and PM state far away from $T_\mathrm{tr}$ to precisely evaluate the changes in the VDOS during the magneto-structural phase transition. For temperature control a LakeShore 340 temperature controller with PID regulation was used with either a silicon diode temperature sensor or a K-type thermocouple, providing a temperature accuracy of $\pm\unit[0.1]{K}$. The Fe-partial VDOS was extracted from the NRIXS data using the Pi program \cite{pi} with correction for residual $\alpha$-Fe contents mentioned earlier. As the ratio of crystallographic abundance of Fe$_\mathrm{I}$ (8b) to Fe$_\mathrm{II}$ (96i) sites in the $\mathrm{La(Fe,Si)_{13}}$ compound is 1:12, the measured Fe VDOS is typical for the dominant Fe$_\mathrm{II}$ species.

In our parameter-free first-principles calculations, we added 3 H-ions to the (24d) positions of the
28 atom
primitive cell of the hydrogen-free compound described earlier \cite{Gruner_2015,cn:Gruner18}, which has three Si atoms placed on
the (96i) sites of Fe. The cell thus corresponds to two formula units (f.u.)
of LaFe$_{11.5}$Si$_{1.5}$H$_{1.5}$ which is a good approximation of the sample stoichiometry used in the experiments. According to Rosca {\em et al.} \cite{Rosca_2010}, we selected only those sites that do not have Si atoms on neighboring (96i) positions, which is one half of the 6 available (24d) sites in the primitive cell. Indeed, our calculations show that this configuration is lower in energy by 316\,meV/H in comparison with the exclusive occupation of the remaining sites, which do have Si in their nearest neighborhood. The calculations were carried out with the Vienna Ab-initio Simulation Package (VASP) \cite{cn:VASP1,cn:VASP2}. The setup is similar as for the hydrogen-free case (see Refs.\ \onlinecite{Gruner_2015,cn:Gruner18} for further details). Exchange and correlation was described with the PW91 functional of Perdew and Wang \cite{cn:Perdew91,cn:Perdew96a} in combination with the spin interpolation formula of Vosko, Wilk and Nusair \cite{cn:Vosko80} and a cutoff energy of $E_\mathrm{cut}$$\,=\,$$380$\,eV. Structural optimizations were carried out on a $k$-mesh of 9$\times$9$\times$9 grid, which yields 125 $k$-points in the irreducible Brillouin zone (IBZ), while for the electronic density of states (DOS) we used a 15$\times$15$\times$15 $k$-grid. For the final presentation, the DOS was convoluted with a Gaussian ($\sigma$$\,=\,$$0.1\,$eV). The calculation of the vibrational density of states (VDOS) is based on the so-called direct or force-constant approach \cite{cn:Kresse95,cn:Gonze97,cn:Parlinski97}. It was calculated with the {\sc PHON} code by Dario Alf\`e{} \cite{cn:Alfe09PHON} using the forces obtained with VASP from 62 individual displacements of $0.02\,{\rm\AA}$ of the inequivalent ions in a 2$\times$2$\times$2 (248 atom) supercell. The paramagnetic (PM) state was represented by a static pseudo-disordered, nearly antiferromagnetic spin-configuration as for the hydrogen-free case \cite{Gruner_2015,cn:Gruner18}, which was stabilized by a fixed-spin-moment constraint \cite{Williams84,Schwarz84,Dederichs84,Moruzzi86} to retain a residual magnetic moment of $3.75\,\mu_{\rm B}$/f.u., similar to Ref.\ \onlinecite{Gruner_2015}.
{\section{\label{sec:results}Results and Discussion}}
\subsection{\label{sec:nrixs}NRIXS and VDOS}
\begin{figure}
    \centering
    \includegraphics[width=0.45\textwidth]{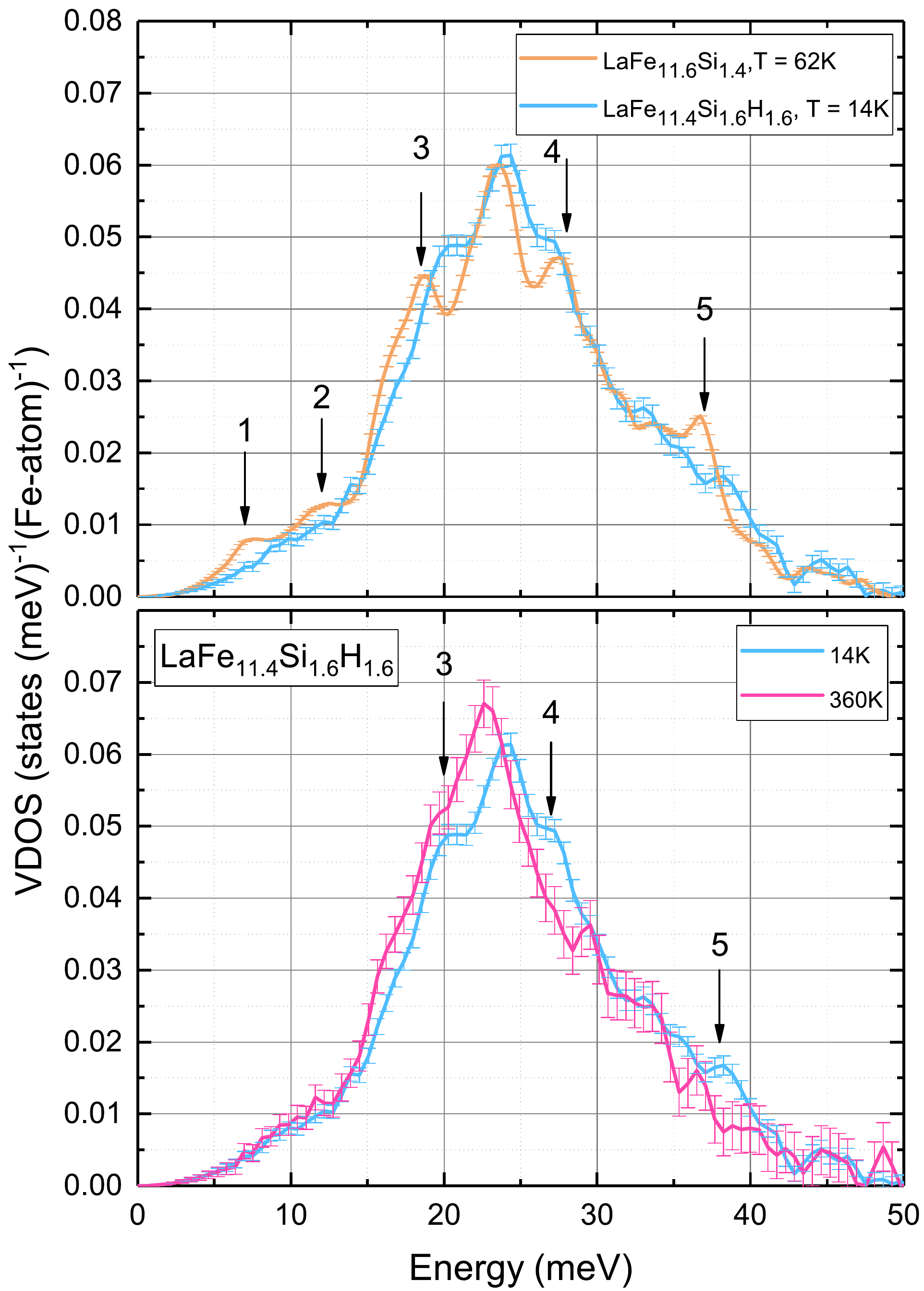}
    \caption{Top: Comparison of the Fe-partial VDOS of $\mathrm{La}\mathrm{Fe}_{11.6}\mathrm{Si}_{1.4}$ (reference sample, orange line) and hydrogenated $\mathrm{La}\mathrm{Fe}_\mathrm{11.4}\mathrm{Si}_\mathrm{1.6}\mathrm{H}_{1.6}$ (blue line). The VDOS were derived from corresponding NRIXS spectra taken at low temperatures, i.e. at $\unit[62]{K}$ for $\mathrm{La}\mathrm{Fe}_{11.6}\mathrm{Si}_{1.4}$ (data taken from \cite{Gruner_2015}) and at $\unit[14]{K}$ for $\mathrm{La}\mathrm{Fe}_\mathrm{11.4}\mathrm{Si}_\mathrm{1.6}\mathrm{H}_{1.6}$, in the FM phase. The black arrows, labeled 1-5 at phonon energies $E$ of $\unit[7,12,18,28]{}$ and $\unit[36]{meV}$, respectively, depict the energetic positions, where sharp peaks exist for the non-hydrogenated samples, and where remarkable changes in the VDOS after hydrogenation are apparent. Bottom: Comparison of the Fe-partial VDOS of the $\mathrm{La}\mathrm{Fe}_\mathrm{11.4}\mathrm{Si}_\mathrm{1.6}\mathrm{H}_{1.6}$ sample in its FM phase at $\unit[14]{K}$ (blue line) and in its PM phase at $\unit[360]{K}$ (pink line). The black arrows at energetic values of $\unit[20]{}$, $\unit[27]{}$ and $\unit[38]{meV}$ depict the change in the VDOS, which occurs after undergoing the metamagnetic phase transition at $T_\mathrm{tr}=\unit[329]{K}$. All data were measured in zero external field. The VDOS with hydrogenation was obtained from sample A with correction of an $\alpha$-Fe content of $\unit[1.91]{\%}$.}
    \label{fig:FM04198_FM_2016}
\end{figure}

    The top graph in Fig.\ \ref{fig:FM04198_FM_2016} shows the (partial) Fe VDOS of non-hydrogenated undoped and hydrogenated $\mathrm{LaFe}_\mathrm{13-x}\mathrm{Si}_\mathrm{x}$ compounds obtained by $^{57}$Fe NRIXS. The $\mathrm{La}\mathrm{Fe}_{11.6}\mathrm{Si}_{1.4}$ sample (reference sample) is identical to the specimen studied in \cite{Gruner_2015} and has been selected for comparison with our present hydrogenated sample that is close to the reference sample in Fe and Si content. Representative NRIXS spectra (raw data) and normalized excitation probability spectra are presented in the supplemental material, \cite{supp}. The spectra depicted in Fig.\ \ref{fig:FM04198_FM_2016} are both taken at low temperature at zero external field in the FM phase, i.e., at $T=\unit[14]{K}$ ($\mathrm{La}\mathrm{Fe}_\mathrm{11.4}\mathrm{Si}_\mathrm{1.6}\mathrm{H}_{1.6}$) and $T=\unit[62]{K}$ ($\mathrm{La}\mathrm{Fe}_{11.6}\mathrm{Si}_{1.4}$). The low temperature (partial) VDOS of $\mathrm{La}\mathrm{Fe}_{11.6}\mathrm{Si}_{1.4}$ exhibits a dominant peak at $\sim\unit[23.5]{meV}$ with weaker peaks 3 and 4 at $\sim\unit[18]{}$ and $\sim\unit[28]{meV}$, respectively. Moreover, there are weak low-energy peaks near $\sim\unit[7]{meV}$ and $\sim\unit[12]{meV}$, numbered 1 and 2, respectively, and a weak high-energy peak at $\sim\unit[37]{meV}$.
    
We observe striking differences between the Fe-partial vibrational density of states with (blue line) and without hydrogenation (orange line) for the FM state (Fig.\ \ref{fig:FM04198_FM_2016} top). In the hydrogenated sample the shape of the VDOS drastically changes with respect to the undoped sample, and the phonon peaks are broadened and shifted. 
Especially the features at about $\unit[7,12,18.5,28]{}$ and $\unit[37]{meV}$, which are marked by the black arrows, are affected. Peaks 1 and 2 at $\unit[7]{}$ and $\unit[12]{meV}$ are reduced to a very low amplitude and are smoothed out by hydrogenation, while peaks 3, 4 and 5 at $\unit[18.5]{meV}$, $\unit[28]{meV}$ and $\unit[37]{meV}$ are reduced to broad shoulders and energetically redistributed. We observe an energetic blueshift of peak 3 (near $\unit[18]{meV}$), a slight blueshift of the main peak (near $\unit[24]{meV}$), a slight redshift for peak 4 (near $\unit[27]{meV}$) and a suppression of peak 5 (near $\unit[37]{meV}$), all of these modifications being induced by hydrogenation in the FM phase (blueshift: shift to higher phonon energies E, redshift: shift to lower phonon energies).
%end attachment D

\begin{figure}
    \centering
    \includegraphics[width=0.45\textwidth]{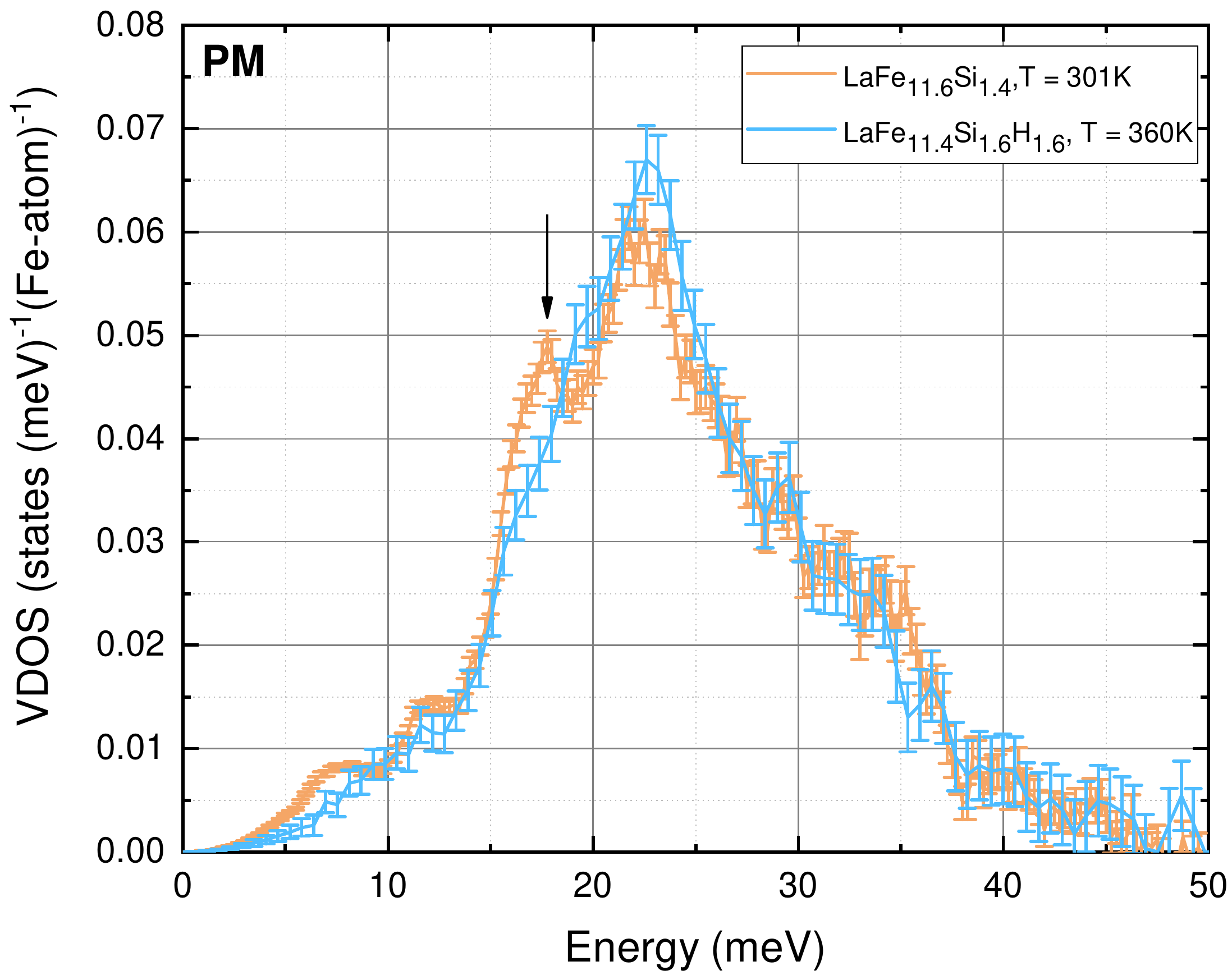}
    \caption{Fe-partial VDOS of $\mathrm{La}\mathrm{Fe}_{11.6}\mathrm{Si}_{1.4}$ at $\unit[301]{K}$ (orange, from Ref. \cite{cn:Landers18}) and $\mathrm{La}\mathrm{Fe}_\mathrm{11.4}\mathrm{Si}_\mathrm{1.6}\mathrm{H}_{1.6}$ at $\unit[360]{K}$ (blue, present work, sample A), both samples being in the PM state. The phonon mode at $\sim\unit[18]{meV}$ (marked by the arrow) of the non-hydrogenated compound is strongly suppressed in the H containing compound. Hydrogenation leads to an overall energetic blueshift of the VDOS in the PM state.} 
     \label{fig:FM04198_PM_2016}
\end{figure}

    To see an imprint of the phase transition onto the vibrational density of states, the VDOS in the two magnetic states have been compared. The Fe-partial VDOS in the low temperature FM phase at $T=\unit[14]{K}$ ($T\ll{}T_\mathrm{tr}$, blue line) and in the high temperature PM phase at $T=\unit[360]{K}$ ($T>T_\mathrm{tr}$, pink line) of hydrogenated $\mathrm{La}\mathrm{Fe}_\mathrm{11.4}\mathrm{Si}_\mathrm{1.6}\mathrm{H}_{1.6}$ in Fig.\ \ref{fig:FM04198_FM_2016} (bottom) reveal clear differences. It can be seen that the residual phonon peaks (shoulders) in the VDOS of the FM state (black arrows, numbered 3-5) at $\unit[20]{}$, $\unit[27]{}$ and $\unit[38]{meV}$ are modified. Peak 3 and 5 are nearly completely quenched in the PM state. In particular the phonon mode 4 visible at $\unit[27]{meV}$ in the FM state is drastically suppressed in the PM state and smoothened out. This is a microscopic signature of strong magnetoelastic (spin-phonon) coupling of the Fe magnetic moments in the FM phase in hydrogenated $\mathrm{La}\mathrm{Fe}_\mathrm{11.4}\mathrm{Si}_\mathrm{1.6}\mathrm{H}_{1.6}$, thus exhibiting similar behavior as observed in our previous studies on non-hydrogenated $\mathrm{LaFe}_\mathrm{13-x}\mathrm{Si}_\mathrm{x}$ \cite{Gruner_2015,cn:Landers18}. The peak reduction in the PM state in the hydrogenated sample is less pronounced than in non-hydrogenated compounds, but the phonon peak at $\sim\unit[27]{meV}$ still vanishes as soon as large magnetic disorder is introduced into the sample. 
 %attachment E
 A similar effect occurs for the $\sim\unit[18]{meV}$ peak (3), which survives the FM-to-PM transition in the non-hydrogenated material \cite{cn:Landers18}, but is strongly suppressed in the PM state of the hydrogenated compound, where it appears as a blueshifted shoulder at $\sim\unit[20]{meV}$ (see Fig.\ \ref{fig:FM04198_PM_2016}).
 %end attachment E   
 
Thus, magnetic disorder occurs upon heating above the phase transition temperature $T_\mathrm{tr}=\unit[329]{K}$ and the formation of the paramagnetic state, leading to distinct modifications in the VDOS.

Most importantly, comparing the magnetically ordered (FM) and disordered (PM) states, a uniform shift of the phonon modes to lower phonon energies (redshift) is visible in the VDOS of Fig.\ \ref{fig:FM04198_FM_2016}, bottom, which, as will be shown below, yields an overall increase in the vibrational entropy.
  The total redshift, which can be seen in Fig.\ \ref{fig:FM04198_FM_2016} (bottom), can be quantified
  in terms of the first ($n=1$) moment of the VDOS $g(E)$,

  \begin{equation}
    \langle E^n\rangle = \int_0^{\infty} E^n \, g(E)\, dE \quad ,
  \end{equation}
  where we assume the integral over $g(E)$ to be normalized to one.
We would like to emphasize that, according to this definition,
$\langle E^1 \rangle$ is distinct from the phonon inner energy $U(T)$ as it
    does not include the Bose-Einstein factor and, therefore, does not include the temperature-dependent phonon occupation probability. $\langle E^1\rangle$ as a function of temperature for $\mathrm{La}\mathrm{Fe}_\mathrm{11.4}\mathrm{Si}_\mathrm{1.6}\mathrm{H}_{1.6}$ and for the reference sample $\mathrm{La}\mathrm{Fe}_{11.6}\mathrm{Si}_{1.4}$ is displayed in Fig. S 10 in Ref.~\cite{supp}. The decrease of
    $\langle E^1\rangle$ amounts to an energetic redshift of
    $\unit[-3.1]{\%}$
    upon heating over the entire measured temperature range from $\unit[14]{K}$ to $\unit[360]{K}$ for $\mathrm{La}\mathrm{Fe}_\mathrm{11.4}\mathrm{Si}_\mathrm{1.6}\mathrm{H}_{1.6}$, while this redshift is
    $\unit[-3.7]{\%}$
    in the measured range from 62 K to 301 K for non-hydrogenated $\mathrm{La}\mathrm{Fe}_{11.6}\mathrm{Si}_{1.4}$ (see Fig. S 10 in Ref.~\cite{supp}). Interestingly, in case of the non-hydrogenated reference sample, a sharp drop of
    $\langle E^1\rangle$ by
    $\unit[-2.6]{\%}$ (redshift) is observed at $T\mathrm{_{tr}}$ = $\unit[192]{K}$ upon heating, with an estimated width of the transition of ~ 20 K (see Fig. S 10). For $\mathrm{La}\mathrm{Fe}_\mathrm{11.4}\mathrm{Si}_\mathrm{1.6}\mathrm{H}_{1.6}$, the drop in
    $\langle E^1\rangle$ at $T\mathrm{_{tr}}$ = $\unit[329]{K}$ (redshift) is reduced to
    $\unit[-1.3]{\%}$,
    which is about half of that of the non-hydrogenated compound, while the width of the transition is much larger ($\sim\unit[50]{K}$) than that of the non-hydrogenated material.
 
\begin{figure}
  \centering
   \includegraphics[width=0.45\textwidth]{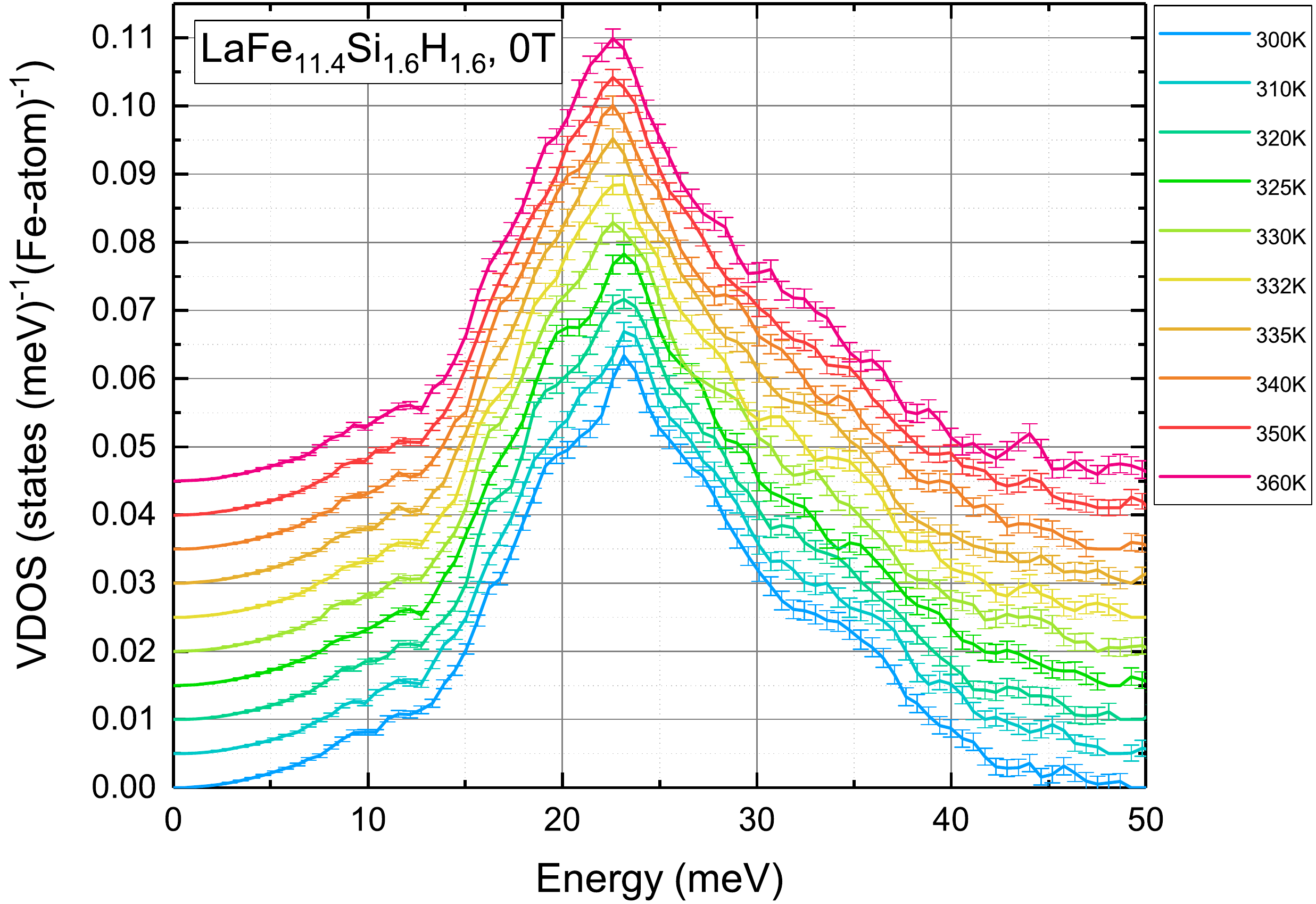}
   \caption{Fe-partial VDOS of $\mathrm{La}\mathrm{Fe}_\mathrm{11.4}\mathrm{Si}_\mathrm{1.6}\mathrm{H}_{1.6}$ measured across the phase transition in zero applied magnetic field and obtained by NRIXS at different temperature points, measured closely around the metamagnetic phase transition, from $\unit[300]{K}$ (bottom line, blue) to above the transition at $\unit[360]{K}$ (top line, pink) following a color gradient. The curves are vertically shifted by 0.005 states $\unit[]{(meV)^{-1} \mathrm{(Fe-atom)}^{-1}}$ for better visualization ($T_\mathrm{tr}\sim\unit[329]{K}$). These VDOS have been corrected for a residual $\alpha$-Fe content of $\unit[4.47]{\%}$ (sample B) and correspond to the entropy points at $\unit[0]{T}$ of Fig.\ \ref{fig:SV}.} 
  \label{fig:FM04198_stack_2017}
\end{figure}

      This redshift (upon heating at the transition temperature) occurs despite the large volume decrease of $\unit[\sim 1]{\%}$, which contradicts the expected behavior following Gr\"uneisen theory \cite{kittel1976, Gruner_2015, cn:Landers18}. From Gr\"uneisen theory, one would expect a blueshift ($\frac{\Delta E}{E}>0$) in phonon energy $E$ (and not a redshift $\frac{\Delta E}{E}< 0$) according to the Gr\"uneisen relation $\frac{\Delta E}{E}= -\gamma(\frac{\Delta V}{V})$, where $\gamma> 0$ is the average Gr\"uneisen constant, and $\frac{\Delta V}{V}<0$ the relative atomic volume contraction at the isostructural FM-to-PM transition at $T_\mathrm{tr}$ for $\mathrm{La}\mathrm{Fe}_\mathrm{11.4}\mathrm{Si}_\mathrm{1.6}\mathrm{H}_{1.6}$. The redshift implies a lattice softening in the PM phase, which has also been observed in previous studies on non-hydrogenated $\mathrm{LaFe}_\mathrm{13-x}\mathrm{Si}_\mathrm{x}$-compounds \cite{Gruner_2015,cn:Landers18}. In the low energy regime ($E<\unit[14]{meV}$), almost no changes in the shape of the VDOS are visible in Fig.\ \ref{fig:FM04198_FM_2016} (bottom). This results from the weak intensity of the phonon modes at low energies.
    Fig.\ \ref{fig:FM04198_stack_2017} shows the VDOS at zero external field of $\mathrm{La}\mathrm{Fe}_\mathrm{11.4}\mathrm{Si}_\mathrm{1.6}\mathrm{H}_{1.6}$ at six temperature points above and four points below the metamagnetic phase transition region, across $T_\mathrm{tr}$. The data corresponds to sample B and has been corrected for a residual $\alpha$-Fe content of $\unit[4.47]{\%}$ as in our former work \cite{Gruner_2015,cn:Landers18}. Significant changes in the VDOS appear closely above and closely below $T_\mathrm{tr}$ = $\unit[329]{K}$ at phonon modes near $\unit[20]{}$ and $\unit[27]{meV}$. These phonon modes decay upon heating across $T_\mathrm{tr}$ and the whole VDOS shifts to lower energies due to the lattice softening with increasing temperature. 
    %attachment E1

    So far we have mostly described the influence of temperature on the VDOS. In the following we will discuss the impact of hydrogenation on the phonon DOS in the low temperature FM state and in the high temperature PM state and the respective overall energetic shifts of the VDOS (see Fig.~S~10 in Ref.\ \cite{supp}). In the low temperature FM state (Fig.\ \ref{fig:FM04198_FM_2016} top) the first moment of the VDOS $\langle E^1\rangle$ of non-hydrogenated $\mathrm{La}\mathrm{Fe}_{11.6}\mathrm{Si}_{1.4}$ at $\unit[62]{K}$ is $\unit[24.68]{meV}$, whereas $\langle E\rangle=\unit[25.22]{meV}$ for hydrogenated $\mathrm{La}\mathrm{Fe}_\mathrm{11.4}\mathrm{Si}_\mathrm{1.6}\mathrm{H}_{1.6}$ at $\unit[14]{K}$. We may neglect the tiny overall redshift of about $\unit[0.1]{\%}$ expected between $\unit[14]{K}$ and $\unit[62]{K}$ upon warming. Then, the influence of hydrogenation on the overall VDOS is a striking energetic blueshift of $\unit[+2.1]{\%}$ in the low-$T$ FM state, as Fig. S 10 in Ref. \cite{supp} demonstrates. Also in the PM state we find an apparent overall blueshift through hydrogenation (see Fig.\ \ref{fig:FM04198_PM_2016}). We end up with a resulting value of $\unit[+3.7]{\%}$ for the blueshift in the PM state of $\mathrm{La}\mathrm{Fe}_{11.4}\mathrm{Si}_{1.4}\mathrm{H}_{1.6}$ at 360\,K, as Fig.~S~10 in Ref.\ \cite{supp} demonstrates. This value is of the same order of magnitude as that in the FM phase. Summarizing, hydrogenation induces an overall blueshift of the Fe-VDOS (phonon hardening) relative to the nonhydrogenated material. This observation agrees with the larger Fe-specific Debye temperature, $\Theta^\mathrm{Fe}$, of the hydrogenated compound (see section \ref{sec_thermo} and Table \ref{tab:Theta}) and indicates lattice hardening
induced
by H atoms.
    
    %end attachment E1
\subsection{\label{sec_thermo}Thermodynamic Properties from NRIXS}

\begin{figure}
    \centering
    \includegraphics[width=0.45\textwidth]{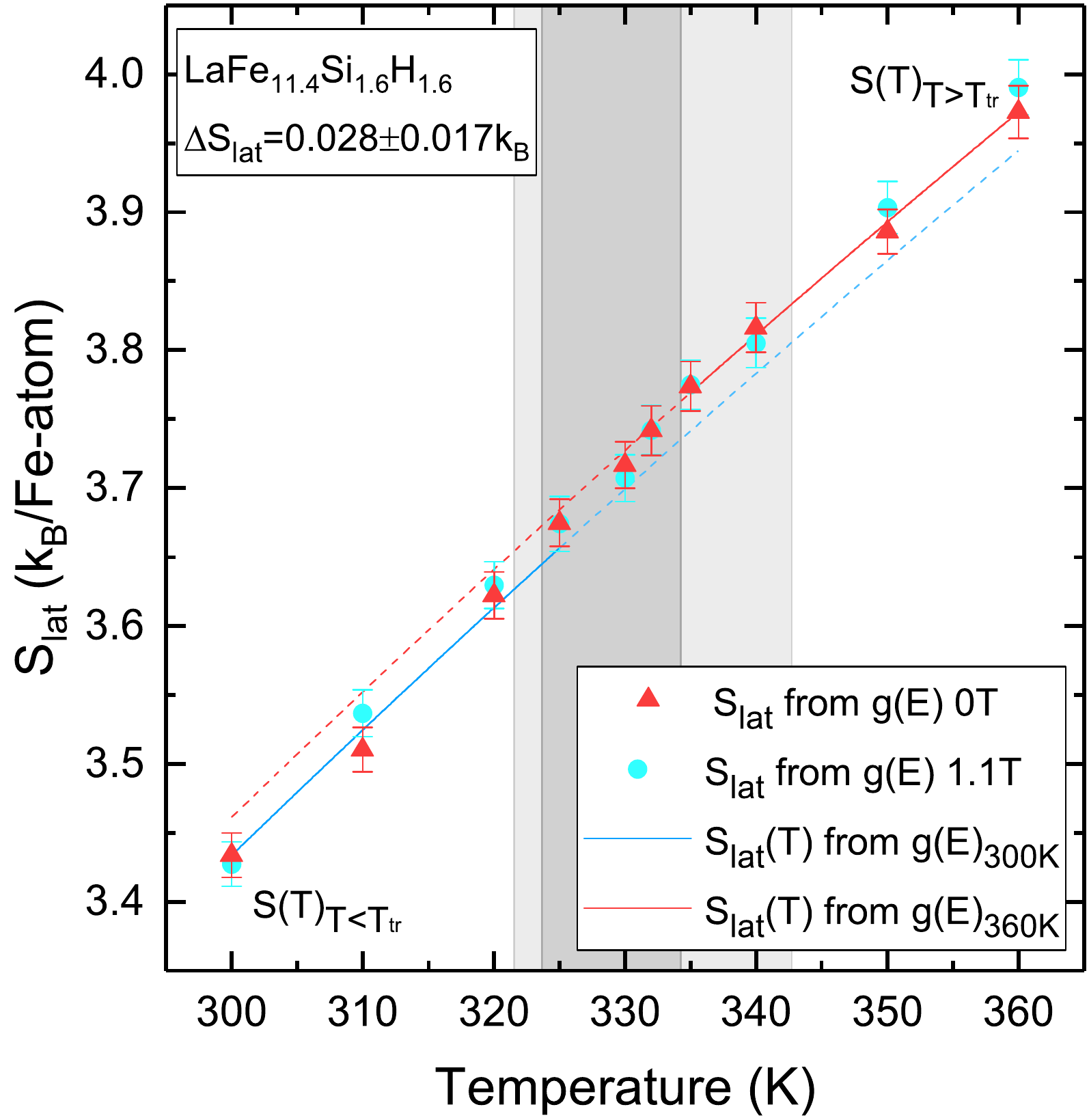}
    \caption{Vibrational (lattice) entropy $S_\mathrm{lat}$, calculated via Eq.\ (\ref{eq:SV}) from the Fe-partial experimental VDOS, $g(E)$, (sample B) at several temperature points across the phase transition. $\mathrm{La}\mathrm{Fe}_\mathrm{11.4}\mathrm{Si}_\mathrm{1.6}\mathrm{H}_{1.6}$ exhibits an increase in $S_\mathrm{lat}$ at the FM-to-PM transition by $\Delta S_\mathrm{lat}=\unit[(0.028 \pm 0.017)]{k_\mathrm{B}}$/Fe-atom. The blue line depicts the calculated $S_\mathrm{lat}(T)$ for the ferromagnetic phase and the red line for the paramagnetic phase, calculated from Eq.\ (\ref{eq:SV}) using the experimentally determined $g(E)$ at the given measurement temperatures, and fitted to the given entropy points in the FM region and PM region, respectively. The data points were taken with rising temperature. Red triangles: from $g(E)$ in zero external field, blue circles: from $g(E)$ in $\mu_0H=\unit[1.1]{T}$. The dark and light gray shaded areas correspond to the phase coexistence region for zero field and $\unit[1.1]{T}$, respectively (compare with Fig.~S~1 in Ref.\ \cite{supp}). The values for $S_\mathrm{lat}$ have been extracted from VDOS (depicted in Fig.\ \ref{fig:FM04198_stack_2017}), which have been corrected for a residual $\alpha$-Fe content of $\unit[4.47]{\%}$ following the method as described in \cite{Gruner_2015,cn:Landers18}. }
    \label{fig:SV}
	\end{figure}

    In order to investigate the thermodynamic behavior of $\mathrm{La}\mathrm{Fe}_\mathrm{11.4}\mathrm{Si}_\mathrm{1.6}\mathrm{H}_{1.6}$, we extracted the (partial) Fe contribution $S_\mathrm{lat}$ to the (total) isothermal entropy $S_\mathrm{iso}$. $S_\mathrm{lat}$ can be directly calculated from the VDOS, $g(E)$, by using the known thermodynamic relation \cite{grimvall1986,cn:Fultz10}
\begin{equation}
S_\mathrm{lat}=3k_\mathrm{B}\int_0^\infty\left(x\coth x-\ln\left(2\sinh x\right)\right)g(E)dE,
\label{eq:SV}
\end{equation}

    with $x=\frac{E}{2k_\mathrm{B}T}$. Fig.\ \ref{fig:SV} provides the vibrational (lattice) entropy $S_\mathrm{lat}$ per Fe atom of hydrogenated $\mathrm{La}\mathrm{Fe}_\mathrm{11.4}\mathrm{Si}_\mathrm{1.6}\mathrm{H}_{1.6}$ for various measurement temperatures across the metamagnetic phase transition, measured with increasing temperatures. The blue line depicts the calculated $S_\mathrm{lat}(T)$ for the ferromagnetic phase from Eq.\ (\ref{eq:SV}), calculated using the experimental VDOS at $\unit[300]{K}$, and the red line shows the same for the paramagnetic phase, calculated from Eq.\ (\ref{eq:SV}), using the experimentally determined $g(E)$ at $\unit[360]{K}$. The temperature points ($\unit[300]{K}$ and $\unit[360]{K}$) have been chosen in order to be in a well defined FM and PM state, yet being as close to the transition temperature as possible, and avoiding the possibility of beginning phase coexistence and a mixture of the two magnetic phases. There is no significant difference between data taken at zero external field and data taken at $\mu_0H=\unit[1.1]{T}$ as the shift in transition temperature is only $\sim\unit[3.5]{\frac{K}{T}}$ and the transition is slightly broadened (see supplement in Ref.\ \cite{supp}). It can be clearly seen that both samples exhibit an offset in the entropy $S_\mathrm{lat}$ after undergoing the phase transition  (the hysteretic region obtained by magnetometry is indicated by the gray shaded area and vertical lines). Upon heating, the increase in the vibrational entropy (related to the redshift of the VDOS at $T_\mathrm{tr}$) in Fig.\ \ref{fig:SV} for the hydrogenated sample has a value $\Delta S_\mathrm{lat}$ of $\unit[(0.028 \pm 0.017)]{k_\mathrm{B}}$/Fe-atom. This corresponds to  $\Delta S_\mathrm{lat}=\unit[(3.2 \pm 1.9)]{\frac{J}{kg K}}$. 
    $\Delta S_\mathrm{lat}$ for the non-hydrogenated reference sample $\mathrm{La}\mathrm{Fe}_{11.6}\mathrm{Si}_{1.4}$ from previous results is $\unit[(0.060\pm 0.023)]{k_\mathrm{B}}$/Fe-atom, or $\unit[(6.9 \pm 2.6)]{\frac{J}{kg K}}$ \cite{cn:Landers18}. The increase in lattice entropy occurs in the metamagnetic phase transition region of $\unit[325-335]{K}$ ($T_\mathrm{tr}=\unit[329]{K}$), where the isostructural phase transition occurs, as derived from our magnetometry measurements. The change in lattice entropy $\Delta S_\mathrm{lat}$ of the Fe subsystem for hydrogenated samples is found to be reduced approximately by half in comparison to non-hydrogenated compounds. 

    From field- and temperature-dependent magnetization measurements, we evaluated a value of $|\Delta S_\mathrm{iso}|=\unit[(9.1\pm0.1)]{\frac{J}{kg K}}$ for the isothermal entropy change from $\unit[0-1.1]{T}$, corresponding to the applied field used in the NRIXS measurements (see Fig. S 3 in Ref \cite{supp}). The obtained entropy change, $\Delta S_\mathrm{lat}$, of $\unit[(0.028 \pm 0.017)]{k_\mathrm{B}}$/Fe-atom or $\unit[(3.2 \pm 1.9)]{\frac{J}{kg K}}$ for only the vibrational contribution of the Fe sub-lattices makes up $\unit[\sim 35]{\%}$ of the total isothermal entropy change and, therefore, strongly contributes to the total entropy.
   
    Furthermore, the entropy Debye temperatures $\Theta_\mathrm{D}$ of the system
    have been calculated from the logarithmic moment of $g(E)$ \cite{grimvall1986,Rosen1983,Eriksson1992,Gruner_2015}:
    \begin{equation}
      k_\mathrm{B}\Theta_\mathrm{D}=\varepsilon\,\mathrm{exp}\left(\frac{1}{3} +
        \int_0^\infty \mathrm{ln}(E/\varepsilon{})\,g(E)\,dE \right).
%        \frac{\int_0^\infty \mathrm{ln}(E/\varepsilon{})g(E)dE}{\int_0^\infty g(E)dE} \right).
        \label{eq:thetadb}
    \end{equation}
    
   Here, $\varepsilon$ is an arbitrary constant carrying the unit of energy.
    
    The entropy Debye temperatures associated with the Fe subsystem
    decrease across the phase transition from the FM to the PM state by roughly
    $\unit[3]{\%}$. We find a value for the entropy Debye temperature for the hydrogenated compound of
    $\Theta_\mathrm{14 K}^\mathrm{Fe}=\unit[(386\pm 7)]{K}$ for the FM phase and a value of
    $\Theta_\mathrm{360 K}^\mathrm{Fe}=\unit[(373\pm 1)]{K}$ in the PM phase (see Tab.\ \ref{tab:Theta}).
    These values are $\unit[\sim15]{K}$ higher than in a non-hydrogenated compound with a nominal
    stoichiometry of $\mathrm{La}\mathrm{Fe}_{11.6}\mathrm{Si}_{1.4}$ \cite{Gruner_2015}.
    This increase of $\Theta^{\mathrm{Fe}}$ of approximately $\unit[4]{\%}$ upon hydrogenation
    is in agreement with the trend in the total Debye temperature $\Theta^{\mathrm{tot}}$ obtained
    from measurements of  the low-temperature specific heat \cite{Lovell2016}.
    Consistent data are also obtained from our first-principles calculations (see Tab. \ref{tab:Theta} in section \ref{sec:DFT}). The agreement in $\Theta^\mathrm{Fe}$ between experiment and theory is excellent.

 \subsection{\label{sec:debyevelocity}Debye velocity of sound from NRIXS}  

\begin{figure}
    \centering
    \includegraphics[width=0.45\textwidth]{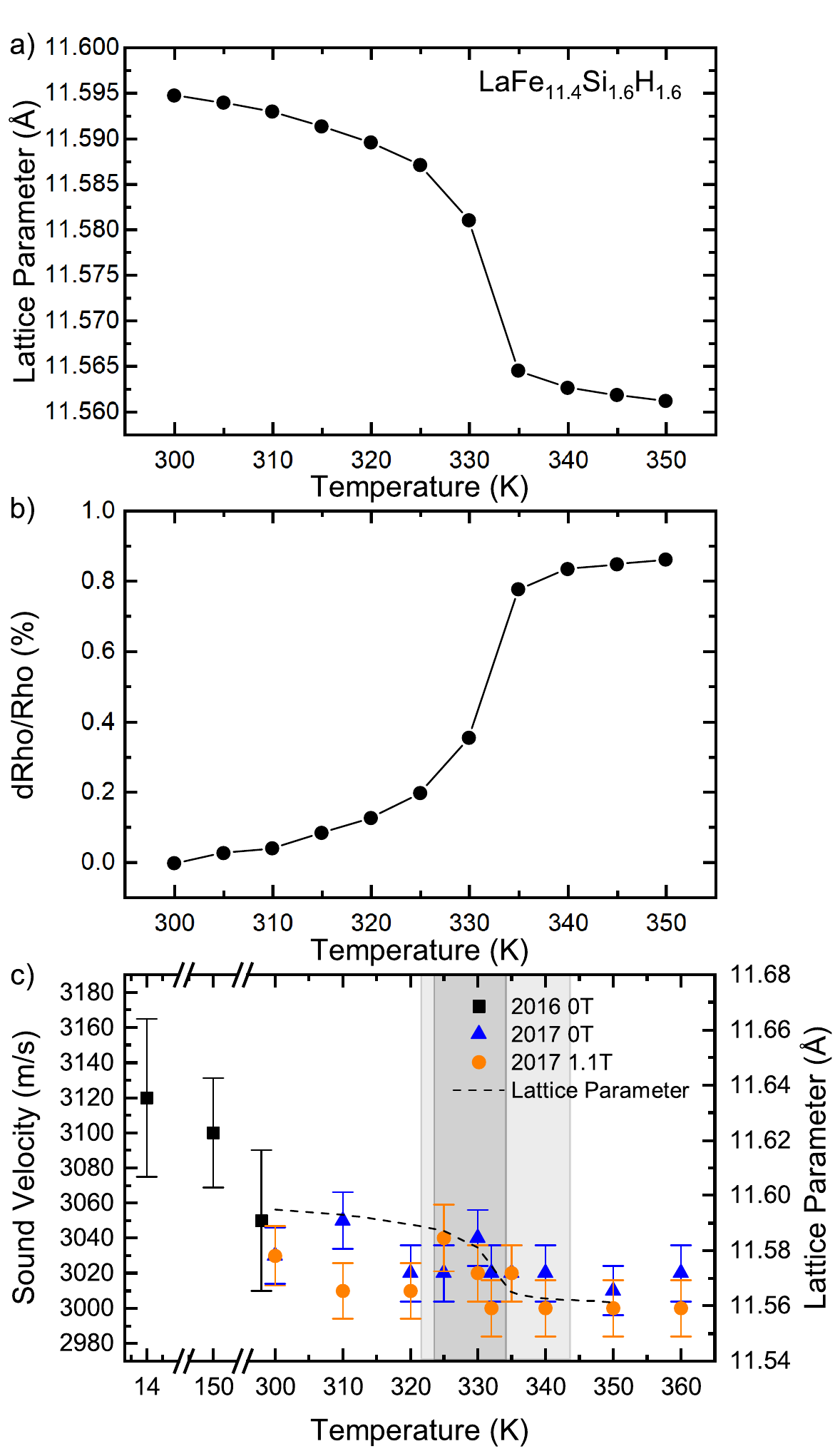}
   \caption{Temperature dependence of the XRD lattice parameter (a), and the relative change of the mass density $\rho$ (b) obtained from (a), for the magnetocaloric hydrogen-containing $\mathrm{La}\mathrm{Fe}_\mathrm{11.4}\mathrm{Si}_\mathrm{1.6}\mathrm{H}_{1.6}$ compound of natural isotopic composition (sample C). ($\rho = \unit[6.992]{\frac{g}{cm^3}}$ at $\unit[300]{K}$). (c) Temperature dependence of the mean sound velocity $\langle v_D\rangle$ (data points) compared to the change in the lattice parameter (dashed line).} 
    \label{fig:vd}
\end{figure}

 The vibrational density of states (VDOS) obtained experimentally from the NRIXS data allows us to determine on the atomic level the average Debye velocity of sound, $\langle v_D\rangle$, in $\mathrm{La}\mathrm{Fe}_\mathrm{11.4}\mathrm{Si}_\mathrm{1.6}\mathrm{H}_{1.6}$ across the magnetostructural phase transition. $\langle v_D\rangle$ has been determined by Herlitschke et al. \cite{Herlitschke2016} employing NRIXS on the magnetocaloric material MnFe$_4$Si$_3$, and by Bessas et al. \cite{Bessas2018} on magnetocaloric (Mn,Fe)$_{1.95}$(P,Si). The velocity of sound is an important physical quantity, which, to the best of our knowledge, surprisingly has not been considered yet in the literature for the important magnetocaloric La(Fe,Si)$_{13}$ and La(Fe,Si)$_{13}$-H compounds. The velocity of sound is a relevant quantity, e.g., for the description of ultrasonic triggering of the giant magnetocaloric effect in thin films \cite{Duquesne2012}.
 Using the VDOS at low energies, $g(E)$,  it is possible to calculate the average Debye velocity of sound, $\langle v_D\rangle$, from the following equation \cite{Achterhold2002,Hu2003a}:
 \begin{equation}
   \lim\limits_{E \rightarrow 0}{\left(\frac{g(E)}{E^2}\right)}=\frac{m_\mathrm{Fe}}{2\pi^2\,\rho\,
     {\langle v_d\rangle^3}\,\hbar^3}
 \label{eq:vd}
 \end{equation}
 The mass of the Fe atom is $m_\mathrm{Fe} = \unit[57\cdot1.66\cdot10^{-27}]{kg}$. $\hbar$ is the reduced Planck constant. $\frac{g(E)}{E^2}$ is the reduced phonon DOS.
 $g(E)$, at low phonon energies $E$, is quadratic in $E$.   
The ratio $\frac{g(E)}{E^2}$ can be described  at low energies by a constant
    called the Debye level \cite{Bessas2018}, which can be determined straight-forwardly from the experimental VDOS
    $g(E)$ \cite{Bessas2018,Herlitschke2016} in the limit $E\to 0$.
$\rho$ in Eq.\ (\ref{eq:vd}) is the mass density of $\mathrm{La}\mathrm{Fe}_\mathrm{11.4}\mathrm{Si}_\mathrm{1.6}\mathrm{H}_{1.6}$, which has not yet been reported across the transition in the literature, to the best of our knowledge. 
Therefore, we have performed temperature-dependent PXRD measurements in the range from $\unit[300]{K}$ to $\unit[350]{K}$ on an independently prepared hydrogen-containing $\mathrm{La}\mathrm{Fe}_\mathrm{11.4}\mathrm{Si}_\mathrm{1.6}\mathrm{H}_{1.6}$ powder sample with natural isotopic composition (sample C, with $T_\mathrm{tr}$ $\sim\unit[332]{K}$, see section II Methods). These PXRD measurements (shown in Fig.\ S~5 of Ref.\ \cite{supp}) provided the lattice parameter (see Fig.\ \ref{fig:vd} (a)) and, consequently, the mass density $\rho$ (see Fig. \ref{fig:vd} (b)) versus temperature across the magnetostructural transition.  The drop in the lattice parameter  observed in the Fig.\ \ref{fig:vd} (a) at the first-order phase transition from the FM to the PM phase amounts to $\unit[0.29]{\%}$, which is in good agreement with the change  observed by neutron diffraction on deuterium-containing $\mathrm{La}\mathrm{Fe}_\mathrm{11.44}\mathrm{Si}_\mathrm{1.56}\mathrm{D}_{1.5}$\cite{cn:Fujieda08}. It can be seen in the Fig.\ \ref{fig:vd} (b) that the mass density increases by $\unit[0.86]{\%}$ when raising the temperature from $\unit[300]{K}$ (below the transition) to $\unit[350]{K}$ (above the transition).

The Debye level has been determined from $g(E)$ in the quadratic low-energy range of $\unit[5.5]{meV} \leq E \leq\unit[10.5]{meV}$ ($g(E)$ at energies smaller than $\sim\unit[5.0]{meV}$ are physically meaningless because of the uncertainty caused by subtraction of the central elastic (M\"ossbauer) peak from the NRIXS spectra, see Fig. S7 in the Supplemental Material \cite{supp}).  The parabolic fit to the $g(E)$ data at low phonon energies below $\sim\unit[10.5]{meV}$ is exemplarily shown for $\mathrm{La}\mathrm{Fe}_\mathrm{11.4}\mathrm{Si}_\mathrm{1.6}\mathrm{H}_{1.6}$ at the lowest ($T = \unit[14]{K}$) and the highest ($T = \unit[360]{K}$) measurement temperature in Fig.~S~9 of the Supplemental Material \cite{supp}. This procedure provides a reasonable estimate of the average Debye velocity of sound for our $^{57}$Fe enriched powder sample (samples A and B). Fig. \ref{fig:vd} (c) displays the average Debye velocity of sound, calculated from the Eq. (\ref{eq:vd}), and its temperature dependence across the phase transition. Due to the large error margins only trends in the $T$-dependence of $\langle v_D\rangle$ can be observed in the Fig. \ref{fig:vd} (c). Above the phase transition at $T_\mathrm{tr}\sim\unit[330]{K} $, $\langle v_D\rangle$ appears to be constant at a value of $(\unit[3020\pm16]){\frac{m}{s}}$ in zero external field (triangular blue symbols), while $\langle v_D\rangle$ in a field of $\unit[1.1]{T}$ appears to be slightly lower at $(\unit[3000\pm16]){\frac{m}{s}}$ (orange full circles). In the range between 300-350K, the data points for $\langle v_D\rangle$ seem to follow the steplike behavior of the lattice parameter (dashed line). Below $T_\mathrm{tr}$, $\langle v_D\rangle$ shows a trend to increase upon cooling both in zero field and applied field, and reaches a zero-field value of $(\unit[3120\pm45]){\frac{m}{s}}$ at $\unit[14]{K}$, which is equal to an increase by $\sim\unit[3]{\%}$ relative to the room-temperature value of about $(\unit[3030\pm16]){\frac{m}{s}}$. The observed overall blueshift of the VDOS upon cooling (Fig. \ref{fig:FM04198_FM_2016}, bottom panel) might contribute to this effect.  An increase of $\langle v_D\rangle$ upon cooling to low temperature has been observed also for magnetocaloric MnFe$_4$Si$_3$ \cite{Herlitschke2016}. We would like to mention that the velocity of sound determined by NRIXS (as shown in the Fig. \ref{fig:vd} (c)) is connected to the THz frequency range of phonons, while ultrasound techniques probe the velocity of sound in the MHz region \cite{Herlitschke2016}. It was shown in Ref. \cite{Herlitschke2016} that the sound velocity of MnFe$_4$Si$_3$ is larger (by $\sim\unit[7]{ \%}$ at room temperature) in the MHz regime than in the THz region. Our values for the sound velocity obtained by NRIXS (Fig. \ref{fig:vd} (c)) are somewhat smaller than those reported in Ref. \cite{Bessas2018} for magnetocaloric (MnFe)$_{1.95}$(P,Si) (FM: $\unit[3661]{\frac{m}{s}}$; PM: $\unit[3267]{\frac{m}{s}}$).

\subsection{\label{sec:DFT}DFT calculations}
\begin{table*}[ht]
  \caption{\label{tab:Theta}
      Entropy Debye temperatures $\Theta$ and sound velocities $\langle v_D\rangle$
      associated  with the Fe-subsystem and the total VDOS obtained from NRIXS measurements and DFT calculations together with the first moment $\langle E^1\rangle$
      of the Fe-partial VDOS.}
\begin{ruledtabular}
\begin{tabular}{ l c c c c c}
\textrm{Temperature}&
\textrm{$\Theta^{\rm Fe}$}&
\textrm{$\Theta^{\rm tot}$}&
$\langle v_D^{\rm Fe}\rangle$ &
$\langle v_D^{\rm tot}\rangle$ &
$\langle E^1\rangle$
\\
\colrule
$\unit[14]{K}$ &  $\unit[386\pm 7]{K}$ & & $\unit[3120 \pm ~~45]{m/s}$ & & $\unit[25.22]{meV}$ \\
$\unit[300]{K}$ & $\unit[377\pm 1]{K}$ & & $\unit[3030 \pm ~~16]{m/s}$ & & $\unit[24.65]{meV}$ \\
$\unit[360]{K}$ & $\unit[373\pm 1]{K}$ & & $\unit[3020 \pm ~~16]{m/s}$ & & $\unit[24.45]{meV}$ \\
\hline\hline
DFT-FM & $\unit[391]{K}$ & $\unit[436]{K}$ & $\unit[3064 \pm ~~69]{m/s}$ & $\unit[3002 \pm 117]{m/s}$ & $\unit[25.60]{meV}$\\
DFT-PM & $\unit[369]{K}$ & $\unit[420]{K}$ & $\unit[2931 \pm 173]{m/s}$ & $\unit[2894 \pm 159]{m/s}$ & $\unit[24.22]{meV}$ \\
\end{tabular}
\end{ruledtabular}
\end{table*}
 
\begin{figure}
    \centering
    \includegraphics[width=0.45\textwidth]{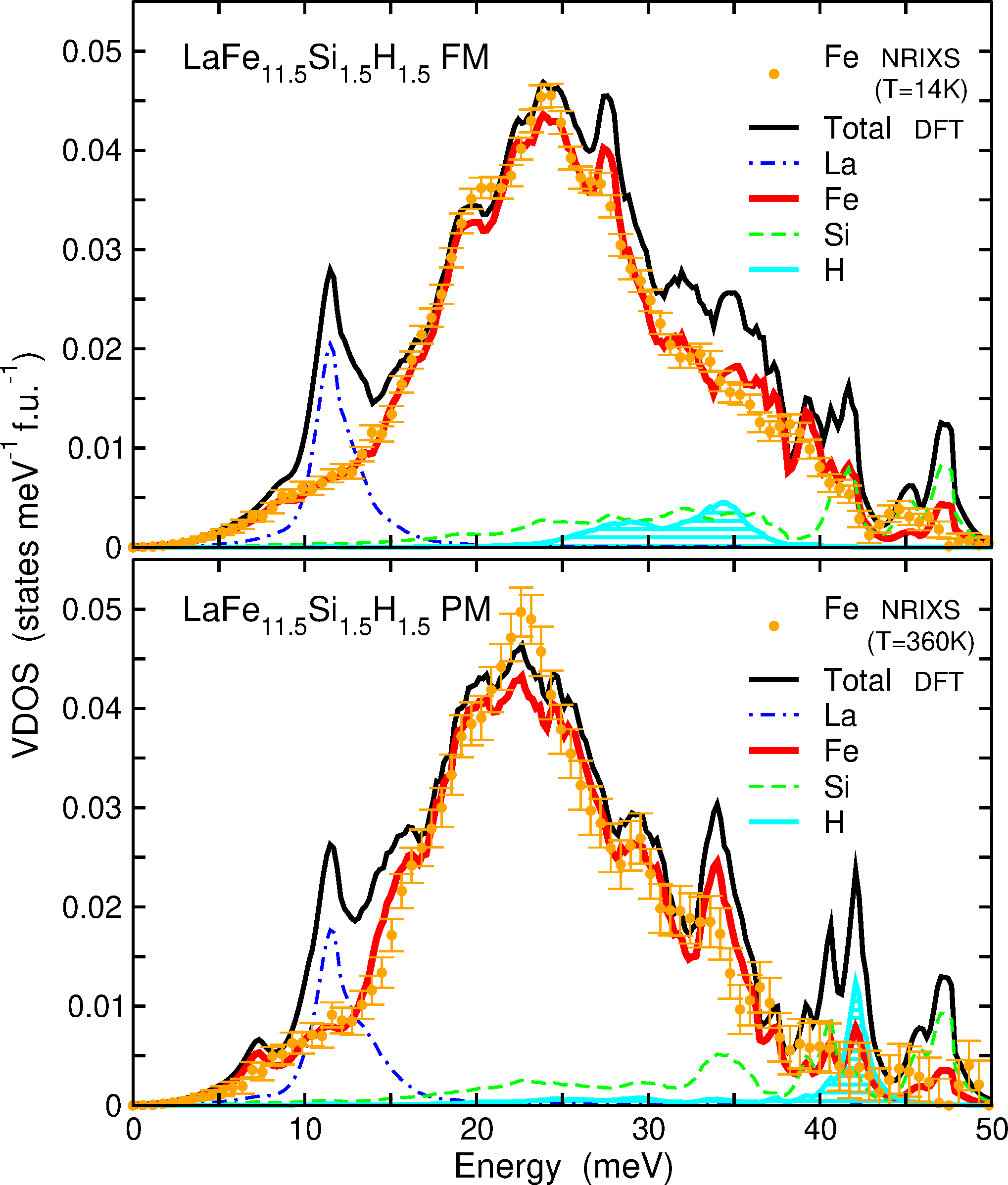}
    \caption{DFT-calculated element-resolved VDOS of hydrogenated $\mathrm{La}\mathrm{Fe}_{11.5}\mathrm{Si}_{1.5}\mathrm{H}_{1.5}$ in the FM (top) and PM phase (bottom).
      Solid black lines denote the total VDOS, colored lines the element resolved
      contributions: Thick red lines for Fe, dash-dotted blue lines for La,
      dashed green lines for Si and solid cyan lines (with hatched area) for H. 
      Orange circles with error bars depict the Fe-partial VDOS obtained
      by NRIXS at $T=\unit[14]{K}$ (FM) and $T=\unit[360]{K}$ (PM). This agrees well with the
      respective Fe contribution from DFT, in particular with respect to the changes in the
      central features discussed in the text. Note, that the energy range does not cover the
      high-lying H-modes located around 150\,meV. %The experimental data has been area-normalized to be compared with the DFT results.
    }
    \label{fig:DFT_FMPM}
\end{figure}
The Fe partial VDOS obtained for the FM and PM states
from first-principle computations agree well with the experimental VDOS measured at low temperatures
($T=\unit[14]{K}$) and at $T=\unit[360]{K}>T_{\rm C}$, as demonstrated 
in Fig.\ \ref{fig:DFT_FMPM}.

This includes Debye temperatures and the average sound velocities $\langle v_D^{\rm Fe}\rangle$ listed in Table \ref{tab:Theta}. The latter were obtained from the average of $g(E)/E^2$ between 2 and 10\,meV.
The DFT values underestimate the experiment consistently by approximately $\unit[80]{m/s}$, which is in the order of the error bars. We consider this an excellent confirmation, as their determination solely involves quantitites
determined from first-principles calculations for the slightly different composition $\mathrm{LaFe}_\mathrm{11.5}\mathrm{Si}_\mathrm{1.5}\mathrm{H}_{1.5}$. More importantly, the difference to the sound velocities obtained
from the total $g(E)$, involving the contribution from the phonon modes from all elements, is of the same magnitude.
This proves that the determination of the average sound velocity from element-selective NRIXS measurements
of the Fe-partial $g(E)$ provides an accurate estimate of this quantity.

DFT predicts a redshift in first moment in $\langle E^1\rangle$ of the $g(E)$ of $-5.4\%$ from the FM
to the PM phase, (see Table \ref{tab:Theta}), which is similar to the experimental value of $-3.1\%$. This trend is similar to the non-hydrogentated case, where we find $\langle E^1\rangle=25.41\,$meV for the FM and $23.00\,$meV for the PM, resulting in a significant redshift of $-9.4\%$, qualitatively consistent with the experimental value of $-3.7\%$ in the range from $\unit[62]{K}$ to $\unit[301]{K}$. Comparing the changes in $\langle E^1\rangle$ upon hydrogenation within the FM and PM structures, we obtain in both cases a blue shift of $+0.7\%$ and $+5.3\%$, respectively, which again reflects the experimental trend ($+2.1\%$ and $+3.7\%$ for the FM and PM phase, respectively, see Fig.~S~10 in Ref.~\cite{supp}.
This proves that the hardening of the Fe-sublattice through hydrogenation is indeed an intrinsic effect,
originating from the incorporation of H on the interstitial (24d) sites.

Even fine details of the experimental VDOS are represented in the calculated VDOS.
Apart from the broadening of the peaks around the central maximum at 18 and 28\,meV,
this includes in particular
the average redshift of the entire DOS and the disappearance of the broadened 28\,meV feature in the PM phase.
For the hydrogen-free case, this was interpreted as an indication of the strong magnetoelastic coupling
present in the system \cite{Gruner_2015,cn:Landers18}.
One third  of the hydrogen vibrational modes are found
at comparatively low energies. This stands in contrast to the general expectation for
light elements, which are supposed to mark the high-frequency end of the vibrational spectrum.
In the present case, the lowest hydrogen modes occupy the same energy range as the 
28-times heavier Si atoms, i.\,e., between 15 and 50\,meV.
According to the low mass, we must expect the exact shape and position of the H-VDOS obtained
from our calculations to be particularly sensitive to the technical settings, including the modeling
of magnetic disorder.
Nevertheless, the low energy of some H-modes indicate a
very shallow direction on the binding surface, where the H atoms can move
almost freely. In other directions this is not the case, since the other two thirds of the vibrational states are distributed between two sharp peaks centered around 150\,meV (omitted in Fig.\ \ref{fig:DFT_FMPM}).
For equiatomic LaFeSiH (tetragonal $P4$/$nmm$ symmetry) it was recently reported that
the vast majority of H modes is seen above 100\,meV, but again a fraction of H modes was found at
low energies around 10\,meV \cite{cn:Hung18}.

\begin{figure}
    \centering
    \includegraphics[width=0.45\textwidth]{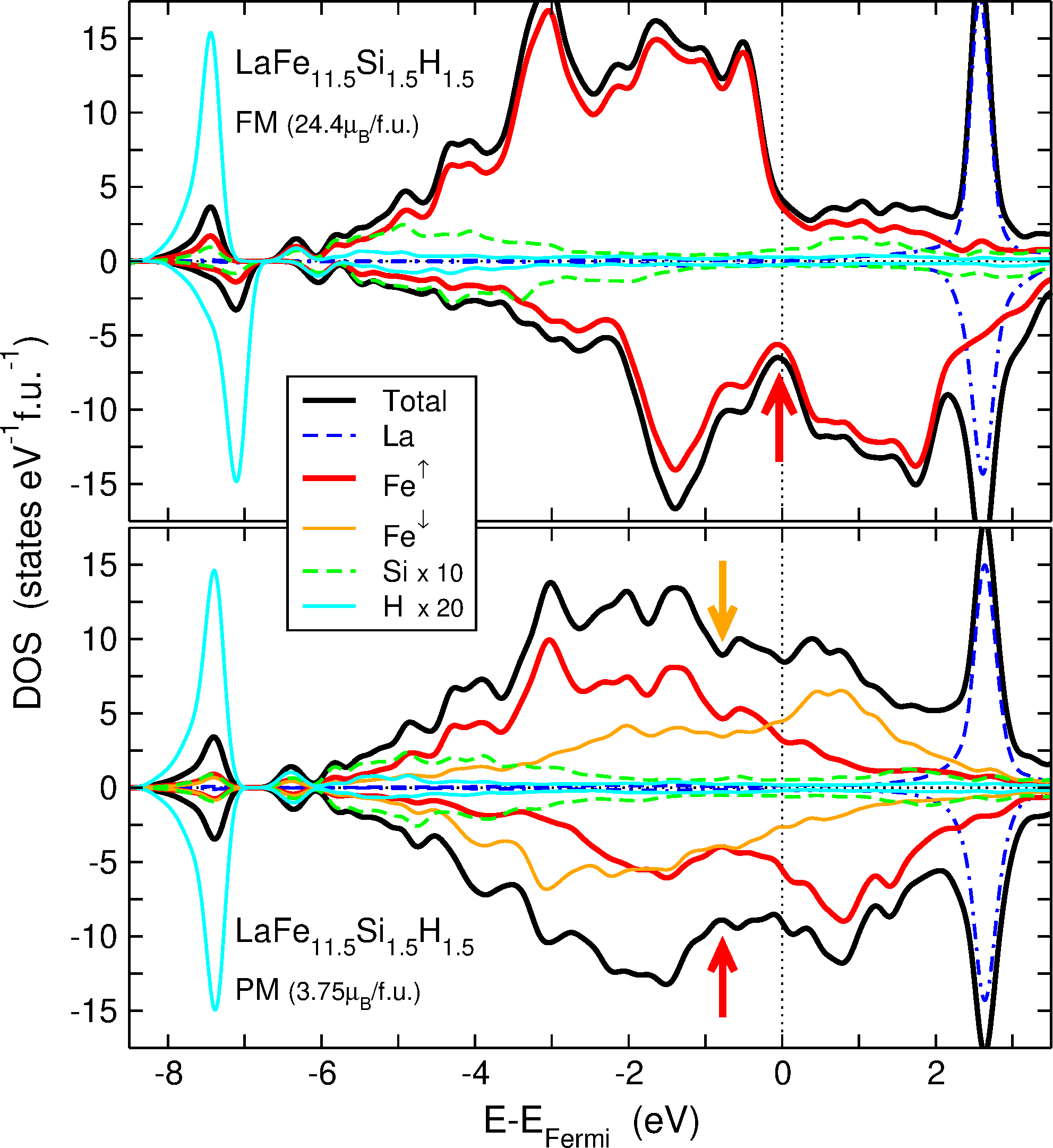}
    \caption{Spin-resolved element- and site-resolved electronic density of states $D(E)$
      of hydrogenated $\mathrm{La}\mathrm{Fe}_{11.5}\mathrm{Si}_{1.5}\mathrm{H}_{1.5}$ in the
      FM (top panel) and PM (bottom panel) phase obtained from DFT. Black lines indicate the total DOS
      of each spin channel, while blue (dash-dotted) and red (solid) lines denote the partial
      DOS of La and Fe$^{\uparrow}$. In the PM state, part of the Fe atoms have opposite magnetic
      orientation (Fe$^{\downarrow}$). Their contribution is indicated by the orange lines.
      For better visibility, their partial contributions of Si (green dashed lines) and
      H (cyan solid lines) are enlarged by a factor of 10 and 20,
      respectively.
      The deep minority spin minimum at the Fermi energy $E_\mathrm{Fermi}$ for the FM state
      is depicted by the vertical red arrow. The red and orange arrows in the bottom panel
      indicate the shift of the minimum in the minority channel
      of the respective site-projected Fe-DOS (Fe$^{\uparrow}$ and Fe$^{\downarrow}$). 
    }
    \label{fig:LaFeSiH_DOS}
\end{figure}
The element-resolved electronic DOS (Fig.\ \ref{fig:LaFeSiH_DOS}) is in close agreement
with the DFT-calculation of Gercsi {\em et al.} \cite{cn:Gercsi18},
which uses a slightly different
setup, where Si replaces Fe on the Fe$_{\rm I}$ (8b)-sites
    instead of the Fe$_{\rm II}$ (96i)-sites.
Fig.\ \ref{fig:LaFeSiH_DOS} reveals a hybridization of
the H- and Fe-states at around -7.5\,eV, which is below the $d$-band edge of Fe
in the hydrogen-free compound.
Such a feature is necessary to explain the stability of the hydrogenated La-Fe-Si,
which corresponds to a gain in formation energy due to the hydrogen uptake. In turn,
there is only a negligible density of states with H-character at the Fermi-level, which
is thus dominated by the $d$-states of Fe in the FM and in the PM phase.
Comparing both phases, we see a picture that is completely analogous
to the hydrogen-free compound. The DOS of the FM phase is characterized by a pronounced
minimum in the minority spin channel right at $E_{\rm Fermi}$
(cf. the arrow in the upper Fig.\ \ref{fig:LaFeSiH_DOS}), which is responsible for
stabilizing the comparatively high magnetic moment of $24.4\,\mu_{\rm B}$/f.u. (or
$2.2\,\mu_{\rm B}$/Fe, respectively), which is likewise
characteristic for the hydrogen-free compound. 

Modeling the PM state
of LaFe$_{11.5}$Si$_{1.5}$H$_{1.5}$ in the same way
as reported in \cite{Gruner_2015,cn:Gruner18} for LaFe$_{11.5}$Si$_{1.5}$ leads once
again to very similar characteristic changes in the DOS, as shown in the lower panel of 
Fig.\ \ref{fig:LaFeSiH_DOS}: The enforced hybridization of minority and majority $d$-states
of Fe-atoms with reversed magnetic orientation leads to the broadening of the features,
including the minimum at $E_{\rm Fermi}$ in combination with a reduction of the local magnetic
moment of Fe to $1.8\,\mu_{\rm B}$/Fe. The result is a decreased exchange splitting,
moving the smeared out minority spin minimum to $0.5-1\,$eV below $E_{\rm Fermi}$,
as indicated by the arrows in the lower panel of Fig.\ \ref{fig:LaFeSiH_DOS}.
The important consequence is again a significantly increased DOS at
$E_{\rm Fermi}$ in the PM phase,
which gives rise to the anomalous softening of the vibrational modes in the PM phase
due to adiabatic electron-phonon coupling \cite{Delaire08,Lucas10,Delaire11,Gruner_2015,Fultz2014}. 
These observations are consistent with earlier temperature-dependent
measurements of the thermopower
\cite{cn:Hannemann12}, which is sensitive to the features in the electronic DOS
close to the Fermi energy. One finds, despite some differences in the
absolute numbers, a rather similar shape of its characteristic change
across the phase transition for the hydrogenated and non hydrogenated compounds.

\begin{figure}
    \centering
    \includegraphics[width=0.45\textwidth]{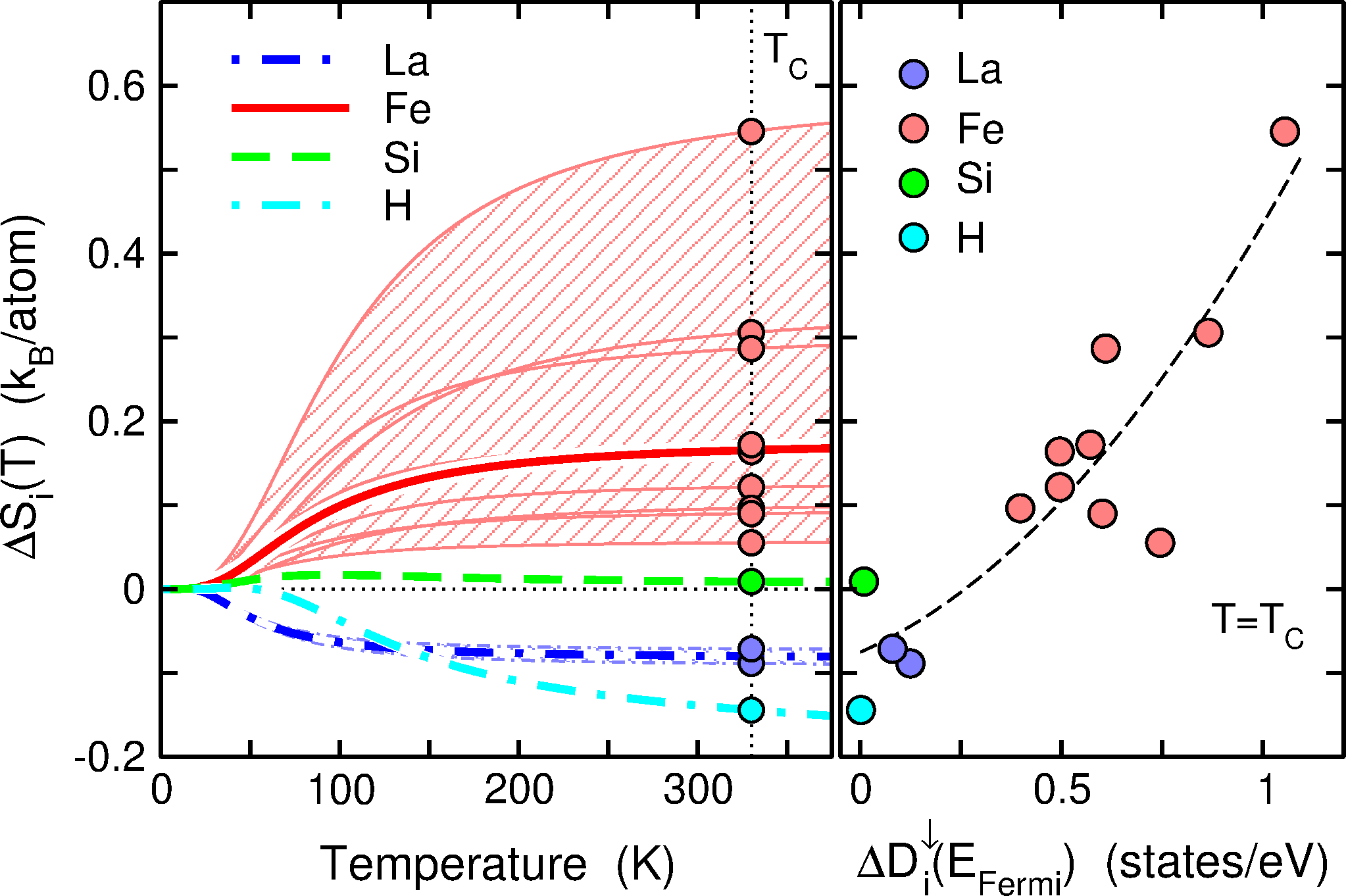}
    \caption{Entropy difference $\Delta S_i$$\,=\,$$S_i(PM)$$\,-\,$$S_i(FM)$
      associated with lattice site $i$ as a function of temperature (left panel)
      calculated from the element- and site-resolved FM- and PM-VDOS obtained by DFT.
      The thick lines denote the element-specific averages.
      In the right panel, the site resolved entropy change (circles)
      at $T_{\rm tr}$$\,=\,$$329\,$K is plotted against the change in the
      site-projected minority spin density of states at the Fermi level
      $\Delta D^{\downarrow}_i(E_{\rm Fermi})$. The dashed line illustrates as a guide to the eye the
      correlation between the two quantities. We interpret this as a consequence of the
      adiabatic electron phonon coupling in hydrogenated La-Fe-Si.
    }
    \label{fig:LaFeSiH_Scorr}
\end{figure}
In analogy to the hydrogen-free case \cite{Gruner_2015},
we can substantiate this picture by comparing
the site-resolved minority spin density of states and the computed
entropy change at the phase transition, as shown in Fig.\ \ref{fig:LaFeSiH_Scorr}.
The thin lines in the left panel show the entropy change as a function of temperature
calculated according to Eq.\ (\ref{eq:SV}) from the respective
site-resolved VDOS in the FM and PM state in Fig.\ \ref{fig:DFT_FMPM}. The thick lines correspond
to the elemental averages. From Fig.\ \ref{fig:LaFeSiH_Scorr} (left) we can directly see, that only the Fe-sites
contribute to the cooperative entropy change at the FM-PM-transition.
The contribution of Si is negligible, whereas La and H
exhibit a decrease in entropy, as expected according to the smaller volume in the PM phase.
Their absolute contribution per site is, however, much smaller in comparison to Fe.
Thus, the presence of H neither affects $\Delta S_{\rm lat}$ nor $\Delta S_{\rm el}$
at $T_{\rm tr}$ significantly.  We ascribe this to the low density
of H-states in the energy range above -6\,eV, where they could hybridize with the Fe-$d$-states,
which are particularly sensitive to the change in magnetic order.

In absolute numbers, the DFT calculations predict
an entropy change $\Delta S_\mathrm{lat}$ related to the Fe sites at $T_{\rm tr}$ of $0.16\,k_{\rm B}$/Fe,
  which is, in analogy to the magnitude of the redshift in $\left< E^1 \right>$, 
significantly larger compared to the experimental
value of $\unit[(0.028 \pm 0.017)]{k_\mathrm{B}}$/Fe-atom reported above.
We ascribe this to the fact that we use idealized models of both phases
in the calculations, while the experimental values were obtained
in close proximity to the phase transition.
Thus, in the calculations, we miss the finite-temperature excitations of the magnetic
subsystem and the associated (negative) thermal expansion in the FM phase.
In turn, in the PM phase, residual FM order (or FM short-range order) may still
be present in the PM phase close to $T_{\rm tr}$. A likewise overestimation of the entropy change
has also been discussed for the non-hydrogenated compound \cite{Gruner_2015,cn:Landers18}.
If we thus compare the relative changes upon hydrogenation, we find
a reduction in $\Delta S_{\rm lat}$ of about 50\,\% after adding hydrogen, consistently
in theory and experiment. This is in agreement with the thermodynamic analysis of
Gercsi {\rm et al.} \cite{cn:Gercsi18}, who concluded on a decrease of
electron-phonon coupling after hydrogenation.

The site-resolved entropy corresponds to the phase space occupied by the relevant
degrees of freedom of a specific ion.
Thus, vibrational entropy can be seen as measure of the elastic properties of the lattice,
since a vibrating ion can occupy more phase space if it moves in a softer potential.
The mechanism of adiabatic electron-phonon coupling links
the availability of states at the Fermi level to the vibrational properties of the
system \cite{Delaire08,Lucas10,Delaire11,Fultz2014}. This has been
identified as the source of the anomalous softening in the hydrogen-free system
\cite{Gruner_2015,cn:Gruner18,cn:Landers18}.
Accordingly, we
expect that also in the hydrogenated compound
the disappearance of the minority spin minimum at $E_{\rm Fermi}$
should lead to a softening
of the lattice, which is consequently reflected in a larger entropy at a
given temperature. The right panel of Fig.\ \ref{fig:LaFeSiH_Scorr} proves that such
a correlation indeed exists, if we compare for each inequivalent lattice site $i$
the entropy change at $T_{\rm tr}$ with the change in the local,
site-resolved minority-spin density of states, taken at the Fermi-level.
\newline
\section{\label{sec:conclusion}Conclusion}
We precisely determined the Fe-partial vibrational (phonon) density of states (VDOS) for hydrogenated $\mathrm{La}\mathrm{Fe}_\mathrm{11.4}\mathrm{Si}_\mathrm{1.6}\mathrm{H}_{1.6}$
by means of $^{57}$Fe NRIXS measurements. We observed characteristic differences in the shape of the VDOS for hydrogenated compounds in comparison to the non-hydrogenated ones, which
are confirmed by first-principles calculations of the element-resolved VDOS
in the framework of density functional theory (DFT).
The temperature evolution of the VDOS across the isostructural phase transition for hydrogenated $\mathrm{La}\mathrm{Fe}_\mathrm{11.4}\mathrm{Si}_\mathrm{1.6}\mathrm{H}_{1.6}$
shows a similar overall trend as observed in non-hydrogenated compounds, but the details are very different.
An overall energetic redshift of the VDOS near $T_\mathrm{tr}$ upon heating can be seen
despite the volume decrease, when undergoing the isostructural first-order phase transition.
%attachment G
Moreover, a striking overall blueshift upon hydrogenation is revealed in the Fe-VDOS for the FM as well as for the PM state (phonon hardening), which is in line with the enhanced Fe-specific Debye temperature observed. Furthermore, from the low-energy part of the experimental Fe-partial VDOS, we determined the average Debye velocity of sound of $\mathrm{La}\mathrm{Fe}_\mathrm{11.4}\mathrm{Si}_\mathrm{1.6}\mathrm{H}_{1.6}$ and its temperature dependence. $\langle v_D\rangle$ is found to be enhanced by $\sim3\,\%$ in the low-T FM state relative to the high-T PM state.
%end attachmentG

Our DFT calculations show, that similar to the hydrogen-free compound, this anomalous behavior must be attributed to  adiabatic electron phonon coupling, which causes the overall lattice softening in the compound at the FM-to-PM transition. It is traced back to the change in the site-resolved minority spin density of states,
originating from a characteristic minimum in the minority spin density of states of the FM right at the Fermi level, which shifts and broadens in the magnetically disordered state. From these results and the reduction of the entropy jump upon heating by $\unit[50]{\%}$ due to hydrogenation, we conclude on a reduction of the adiabatic electron-phonon coupling, confirming the previous conjecture of Gercsi {\rm et al.} \cite{cn:Gercsi18} based on different experimental and theoretical data.

In addition, we observe a strong reduction of prominent phonon modes and a broadening of the phonon peaks upon hydrogenation. The Fe-specific phonon mode at $\sim\unit[27]{meV}$ is reduced to a shoulder in the hydrogenated compound, which disappears after undergoing the phase transition from the FM to the PM state. Interestingly, the phonon mode near $\unit[18]{meV}$, which is retained in the PM state of non-hydrogenated $\mathrm{La}\mathrm{Fe}_\mathrm{11.6}\mathrm{Si}_\mathrm{1.4}$, is strongly quenched in the PM state due to hydrogenation. These effects are another microscopic manifestation of strong magnetoelastic
coupling in $\mathrm{La}\mathrm{Fe}_\mathrm{11.4}\mathrm{Si}_\mathrm{1.6}\mathrm{H}_{1.6}$.
%attachmentG pt2
Our results reveal that hydrogen does not only shift the temperature of the first-order transition, but also significantly affects the magnetoelastic response of the Fe subsystem, as observed experimentally and theoretically. We would like to mention that NRIXS investigations on other magnetocaloric materials \cite{Herlitschke2016,Bessas2018}, although they also demonstrated a clear redshift in the Fe-partial  VDOS across the first-order phase transition, did not reveal any changes in the shape of the Fe-partial VDOS across the transition. Thus, La(Fe,Si)$_{13}$ and La(Fe,Si)$_{13}$H appear to be exceptional materials with respect to strong  spin-phonon coupling and strong adiabatic EPI for particular Fe-specific  phonon modes.
%end attachment G pt 2
%attachment K
The fingerprint of pronounced spin-phonon coupling is the Fe-specific phonon mode near $\unit[27]{meV}$ in both non-hydrogenated and hydrogenated La(Fe,Si)$_{13}$.
  
In this context it is interesting that a remarkable phonon peak at the same energy ($\sim\unit[28]{meV}$) (also related to Fe vibrations) has been predicted for superconducting LaFeSiH (the 1:1:1:1 compound) \cite{cn:Hung18}. This suggests that this Fe phonon mode has the same origin in these materials and experiences strong electron-phonon interaction. Future $^{57}$Fe NRIXS experiments are needed to support this assumption.
%end attachment K

From the VDOS we derived the vibrational (lattice) entropy $S_\mathrm{lat}$ of the
Fe subsystem across the magneto-structural phase transition.
For $\mathrm{La}\mathrm{Fe}_\mathrm{11.4}\mathrm{Si}_\mathrm{1.6}\mathrm{H}_{1.6}$ the vibrational entropy
$S_\mathrm{lat}$, directly extracted from the Fe partial VDOS, $g(E)$, exhibits an increase by
$\unit[(0.028 \pm 0.017)]{k_\mathrm{B}}$/Fe-atom (or $\unit[(3.2 \pm 1.9)]{\frac{J}{kg K}}$)
upon heating across the transition temperature ($T_\mathrm{tr}=\unit[329]{K}$).
This value is found to be only half the value of non-hydrogenated compounds
\cite{cn:Landers18}, which agrees well with the trend
of the values found by first-principles calculations.
It contributes with $\unit[\sim 35]{\%}$ to the overall isothermal entropy
change $\Delta S_\mathrm{iso}$ at $T_\mathrm{tr}$ ($\unit[(9.1\pm0.1)]{\frac{J}{kg K}}$). Accordingly, the entropy Debye temperature
of the Fe subsystem $\Theta_\mathrm{D}^\mathrm{Fe}$,
which turns out to be approximately $\unit[4]{\%}$ larger than in the hydrogen-free case,
exhibits a decrease of $\sim\unit[3]{\%}$ from the FM to the PM state.
%attachment H

Although we have observed distinct modifications in the vibrational density of states (VDOS) and electronic DOS and a $\unit[50]{\%}$ reduction in $\Delta S_\mathrm{lat}$ in hydrogenated La(Fe,Si)$_{13}$H as compared to the non-hydrogenated material,  the total change $\Delta S_\mathrm{iso}$  is known to be almost the same for the two materials at their respective transition temperatures \cite{Fujita2003}. Thus, the $\sim\unit[50]{\%}$ decrease of $\Delta S_\mathrm{lat}$ in the hydrogenated material must be compensated by other terms in equation (\ref{eq:Siso}) above.

Regarding the fact that a hydrogen atom provides an extra electron to the conduction electron density of states, we can look for other elements with such a property. Recently, a huge entropy change of $\Delta S_\mathrm{mag} = \unit[31.4]{ \frac{J}{kgK}}$ at a magnetic field change of only $\unit[3]{T}$ was reported \cite{Jin2019} for the slightly P-doped LaFe$_{11.6}$Si$_{1.4}$P$_{0.03}$ compound near the Curie temperature of $\unit[194]{K}$, with the P atoms most likely occupying 96i sites, like Fe and Si atoms. As P atoms have an outer shell of $3s^23p^3$, P adds one electron more than a Si atom ($2s^22p^2$) to the conduction electron system. In order to check whether this effect contributes to this observed giant $\Delta S_\mathrm{mag}$ enhancement, doping (possibly by non-equilibrium techniques, such as splat cooling, melt spinning or ion implantation) with other elements in the IVa column of the periodic table (As, Sb, Bi) could be of interest.
%end attachment H

Our work proves that, while hydrogenation is capable of shifting the transition to ambient
conditions, the inherent adiabatic electron phonon coupling and large moment-volume
coupling \cite{cn:Gruner18} remain effective. These are associated with the itinerant electron metamagnetism, which determines the characteristic properties of the hydrogen-free system.
Thus the cooperative contribution
of the various degrees of freedom to the magnetocaloric effect persists after
hydrogenation, which
contributes to the superior magnetocaloric performance of this system.

\begin{acknowledgments}The authors would like to thank Dr. M. Krautz (IFW Dresden) for helpful discussions and U. von H\"orsten (University of Duisburg-Essen) for outstanding technical assistance. Funding by the DFG via SPP 1599 (WE2623/12-2, GR3498/3-2, GU514/6) Ferroic Cooling, SPP 1681 (WE2623/7-2), FOR 1509 (WE2623/13-2), CRC 1242 (project A5), (WE2623/14-1) and CRC/TRR 247 (project B2) is gratefully acknowledged. Use of the Advanced Photon Source, an Office of Science User Facility operated for the US Department of Energy (DOE) Office of Science by Argonne National Laboratory, was supported by the US DOE under Contract No. DE-AC02-06CH11357. Part of the calculations were carried out on Cray XT6/m and magnitUDE (DFG grants INST 20876/209-1 FUGG, INST 20876/243-1 FUGG)
supercomputer systems of the Center for Computational Sciences and Simulation (CCSS)
at the University of Duisburg-Essen. 
\end{acknowledgments}
%
%\bibliography{LaFeSiH.bib}
\end{document}